\documentclass[a4paper,11pt]{article}
\pdfoutput=1 % if your are submitting a pdflatex (i.e. if you have
\usepackage{orcidlink}
\usepackage{lineno}
\usepackage{setspace}
\usepackage{verbatim}

\usepackage{standalone}
\usepackage{siunitx}
\usepackage{import}
\usepackage{tikz}
\usepackage{graphicx}
\usepackage{subcaption}
\usepackage[version=4]{mhchem}

\usepackage{multirow}

\usepackage{jinstpub} % for details on the use of the package, please
\usepackage{cleveref}
\title{Characterization of thin optical filters for high purity Cherenkov light readout from scintillating crystals}% 

\newcommand\dnote[2][]{%
\setcounter{footnote}{1}
\renewcommand{\thefootnote}{\fnsymbol{footnote}}
\if!#1!%
\stepcounter{footnote}\footnotetext{#2}%
\else%
{\renewcommand\thefootnote{#1}%
\footnotetext{#2}}%
\fi}

\author[2]{A.~Benaglia,}
\author[2,3]{F.~Cetorelli,}
\author[2,3]{M.~T.~Lucchini,}

\author[1]{E.~Auffray,}
\author[1,4]{L.~Roux,}
\author[1,5]{J.~Delenne}
\affiliation[1]{European Center for Nuclear Research, Geneva, Switzerland}

\affiliation[2]{INFN Sezione di Milano-Bicocca, Milano, Italy}

\affiliation[3]{Università di Milano-Bicocca, Milano, Italy}

\affiliation[4]{Université Claude Bernard Lyon 1, UMR5306 CNRS ILM, Lyon, France}

\affiliation[5]{Université de Strasbourg, UMR7178 CNRS IPHC, Strasbourg, France}

\abstract{A hybrid dual-readout calorimeter concept, comprising both electromagnetic and hadronic sections, has recently been proposed to meet the performance requirements of experiments at future e$^{+}$e$^{-}$
colliders. The front compartment consists of a homogeneous electromagnetic calorimeter made of high-density crystals, each coupled to a pair of Silicon Photomultipliers (SiPMs) providing the simultaneous readout of scintillation and Cherenkov light. To efficiently detect Cherenkov photons in the presence of dominant scintillation signals, an optical filter is placed in front of one of the two SiPMs to suppress photons in the wavelength region corresponding to that of scintillation emission.
In this study, PWO, BGO, and BSO crystals with different dimensions were tested to measure their scintillation light yield and decay time, as well as their transmission and emission spectra. A set of $\sim 100~\rm \mu m$-thick optical filters was also characterized by measuring their transmittance curves. The experimental results were used to model and estimate the expected filter performance in attenuating scintillation light for the various crystals.
The performance of each filter was experimentally validated by measuring the crystal light output with and without the filter using a $^{22}$Na radioactive source and a LYSO:Ce crystal, confirming the accuracy of the calculations.
The results show that interference filters are unsuitable for this application because their transmittance strongly depends on the photon incidence angle. Conversely, two absorptive long-pass filters with cutoff wavelengths around 590~nm were found to block more than 99\% of the scintillation light from PWO crystals, satisfying the calorimeter specifications.
}


\begin{document}

\maketitle

\flushbottom
\section{Introduction}
%New detector concepts are being explored to address the challenges of future collider experiments exploiting recent technological developments.
To continue the exploration of fundamental physics processes, including a detailed characterization of the Higgs boson and searches for hints of new physics beyond the Standard Model, a new e$^{+}$e$^{-}$ collider is being considered by the high energy physics community as one of the best strategy \cite{European:2720131}.
In this context, a novel electromagnetic calorimeter concept consisting of highly segmented scintillating crystals readout with Silicon Photomultipliers (SiPMs) optimized for operation at such a future collider, has been proposed \cite{Lucchini_2020}.
In addition to an electromagnetic energy resolution better than 3\%$/\sqrt{E}$, the detector will feature the simultaneous readout of scintillation (S) and Cherenkov (C) signals. Once combined with a fiber-based dual-readout hadronic calorimeter section \cite{Ampilogov_2023}, this will enable the possibility to use the dual-readout method \cite{RevModPhys.90.025002} to reconstruct hadrons and jets with an energy resolution at the level of $30\%/\sqrt{E}$.

As proposed in \cite{Lucchini_2020}, the implementation of dual-readout in the crystal calorimeter can be achieved by using two separate SiPMs, one with an optical filter to shield the scintillation light and let through only Cherenkov photons with a wavelength different than that of scintillation photons.
To achieve its performance goals the calorimeter design requires a light output of about 2000 scintillation photoelectrons (phe) per GeV of energy deposited in the crystals and about 50 phe/GeV from Cherenkov radiation.
Furthermore, simulation studies indicate that the contamination of scintillation photons to the Cherenkov signal should be smaller than 20\% for an optimal performance of the dual-readout method.

The choice of an adequate optical filter is thus critical to isolate and detect a sufficient number of Cherenkov photons, $N_C$, over the typically much larger number of scintillation photons $N_S$.
Other parameters, such as the crystal and SiPM geometry also play a relevant role in the absolute numbers of photoelectrons detected and are being matter of optimization studies.

We present in this paper, a detailed characterization of a set of optical filters and a measurement of their efficiency in shielding scintillation light from three crystal candidates for this calorimeter concept: namely lead tungstate (PWO), bismuth silicate (BSO) and bismuth germanate (BGO).
A detailed characterization of the scintillation and optical properties of the crystals is discussed in Section~\ref{sec:crystal_char} including a discussion of the role of the crystal and SiPM dimensions on the light output.
The transmittance of the filters and their performance in shielding scintillation light is presented in Section~\ref{sec:filters_trans} and Section~\ref{sec:filter_validation}, respectively. 
The results demonstrate that the fraction of scintillation light passing through a given filter can be estimated from the convolution of the crystal emission spectrum and the filter transmittance. The filter performance estimated in this way is in agreement with the experimental measurements within 20\% as long as the wavelength dependence of the SiPM photon detection efficiency and the angular dependence of the filter transmittance (for interference filters) are taken into account.
In particular, we confirmed that interference filters are not suitable to shield scintillation photons which exit from the crystals with a broad angular distribution spectrum (from 0 to 180$^{\circ}$) and we identified instead a set of 100~$\rm \mu m$ thick absorptive long-pass filters with cut-off wavelength around 590~nm which can filter out more than 99\% of the scintillation light and meet the requirements of the calorimeter discussed in \cite{Lucchini_2020}.

%\newpage
%\clearpage

\section{Crystal characterization}\label{sec:crystal_char}
A set of fully polished BGO, BSO and PWO crystals with different dimensions was manufactured by SICCAS (Shanghai Institute of Ceramics, China) and underwent a detailed optical and scintillation characterization. The crystals featured different lengths and transverse sections, as reported in table \ref{tab:calo_feat_compare}, to provide an evaluation of intrinsic light absorption and light collection efficiency as a function of the crystal dimensions.

\begin{table}[!htbp]
\centering \caption{List of crystals tested.} \vspace{0.2cm}
\begin{tabular}{c|c|c|c}
\hline
Crystal type &    Manufacturer  & Available lengths    & Section \\ 
\hline\hline
BGO          &    SICCAS        & 50~ mm    & $8\times8$~mm$^2$ \\ 
BGO          &    SICCAS        & 50~ mm    & $12\times12$~mm$^2$ \\ 
BGO          &    SICCAS        & 10 / 50 / 130 / 160~ mm    & $10\times10$~mm$^2$ \\ 
\hline
BSO          &    SICCAS        & 10 / 50 / 130~ mm    & $10\times10$~mm$^2$ \\ 
\hline
PWO          &    SICCAS        & 10 / 50 / 130~ mm    & $10\times10$~mm$^2$ \\ 
\hline
\end{tabular} 
\label{tab:calo_feat_compare}
\end{table}
\vspace{0.6cm}

\subsection{Transmittance}
\begin{figure}[!tbp]
    \centering    
    \includegraphics[width=0.329\linewidth]{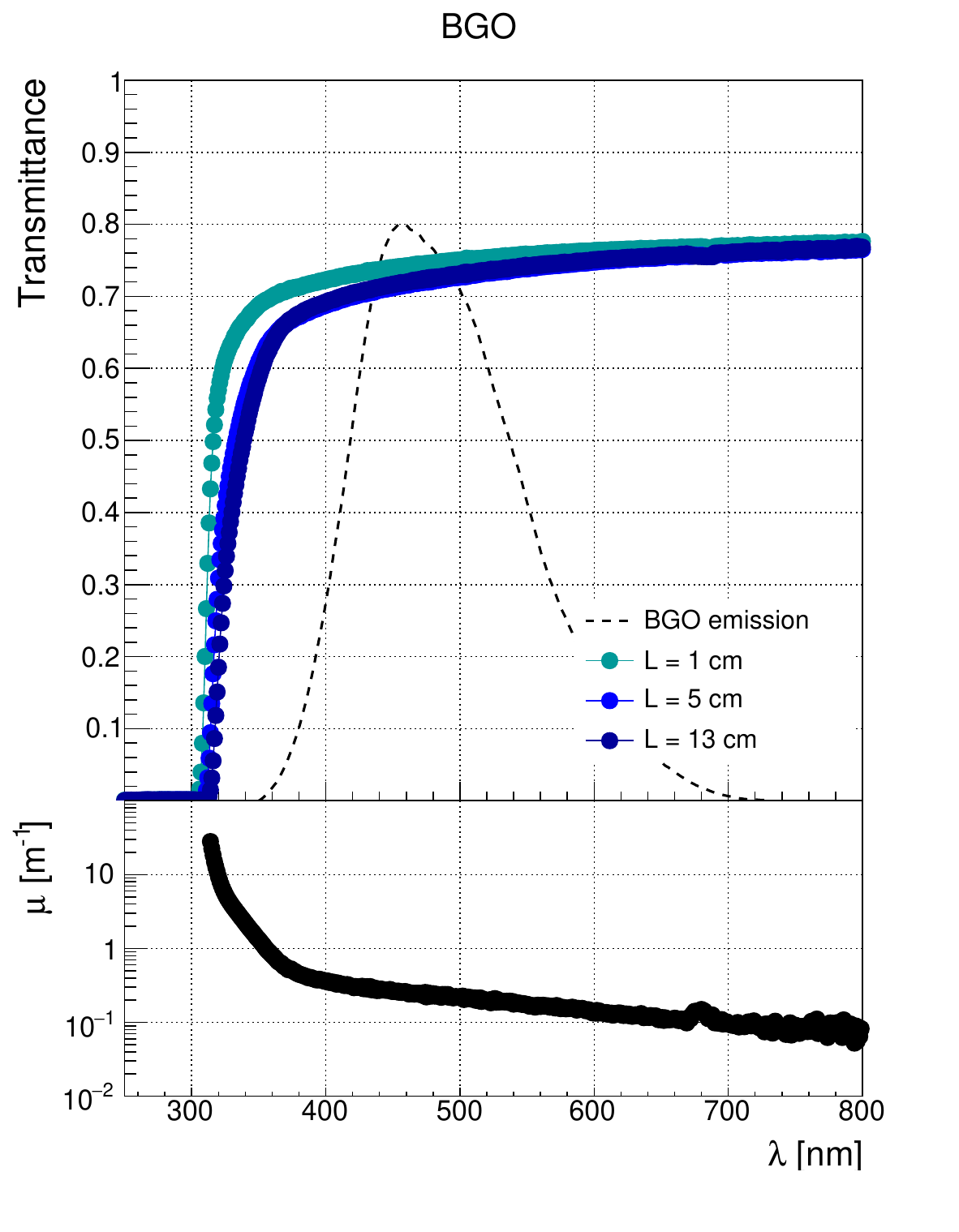}
    \includegraphics[width=0.329\linewidth]{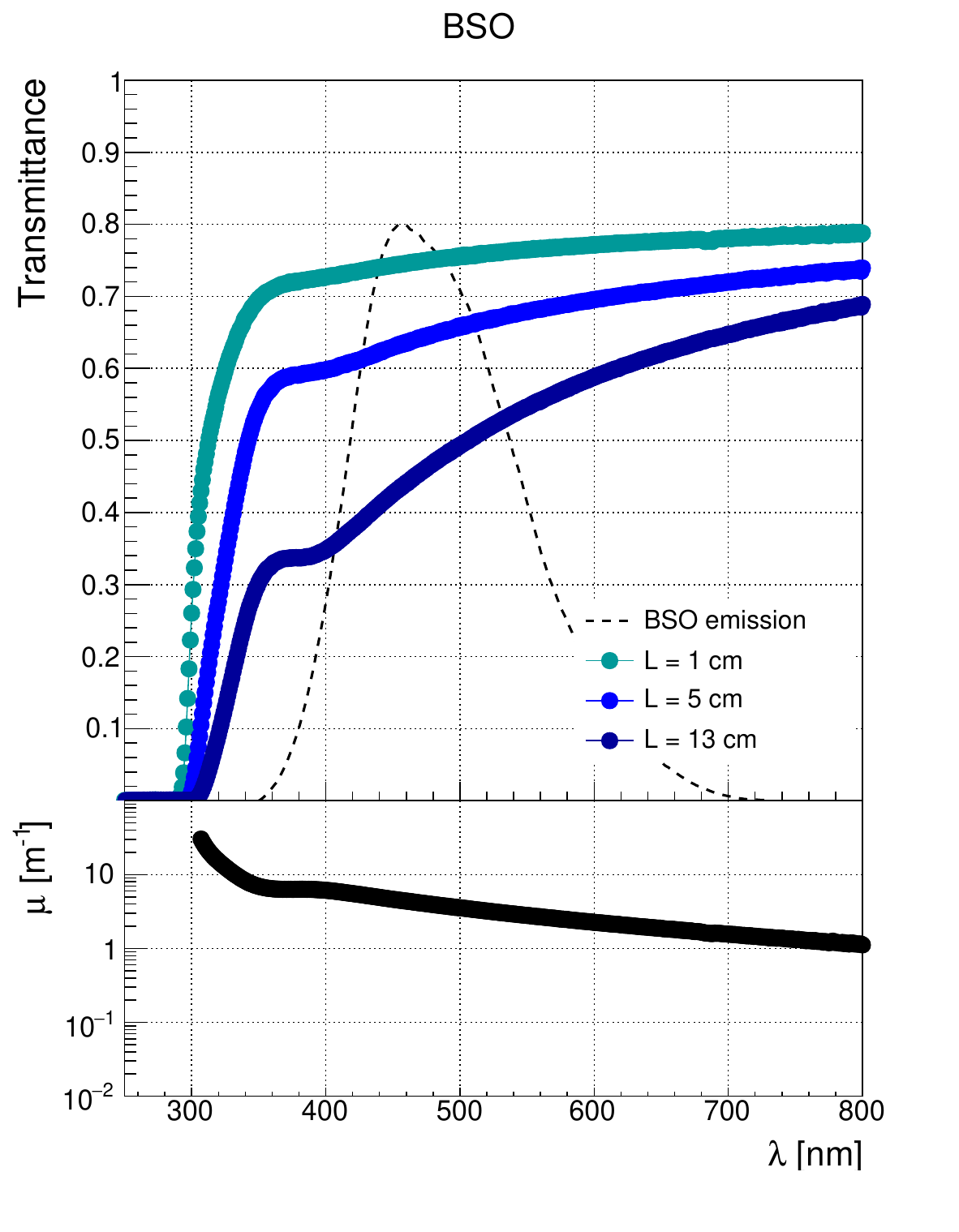}
    \includegraphics[width=0.329\linewidth]{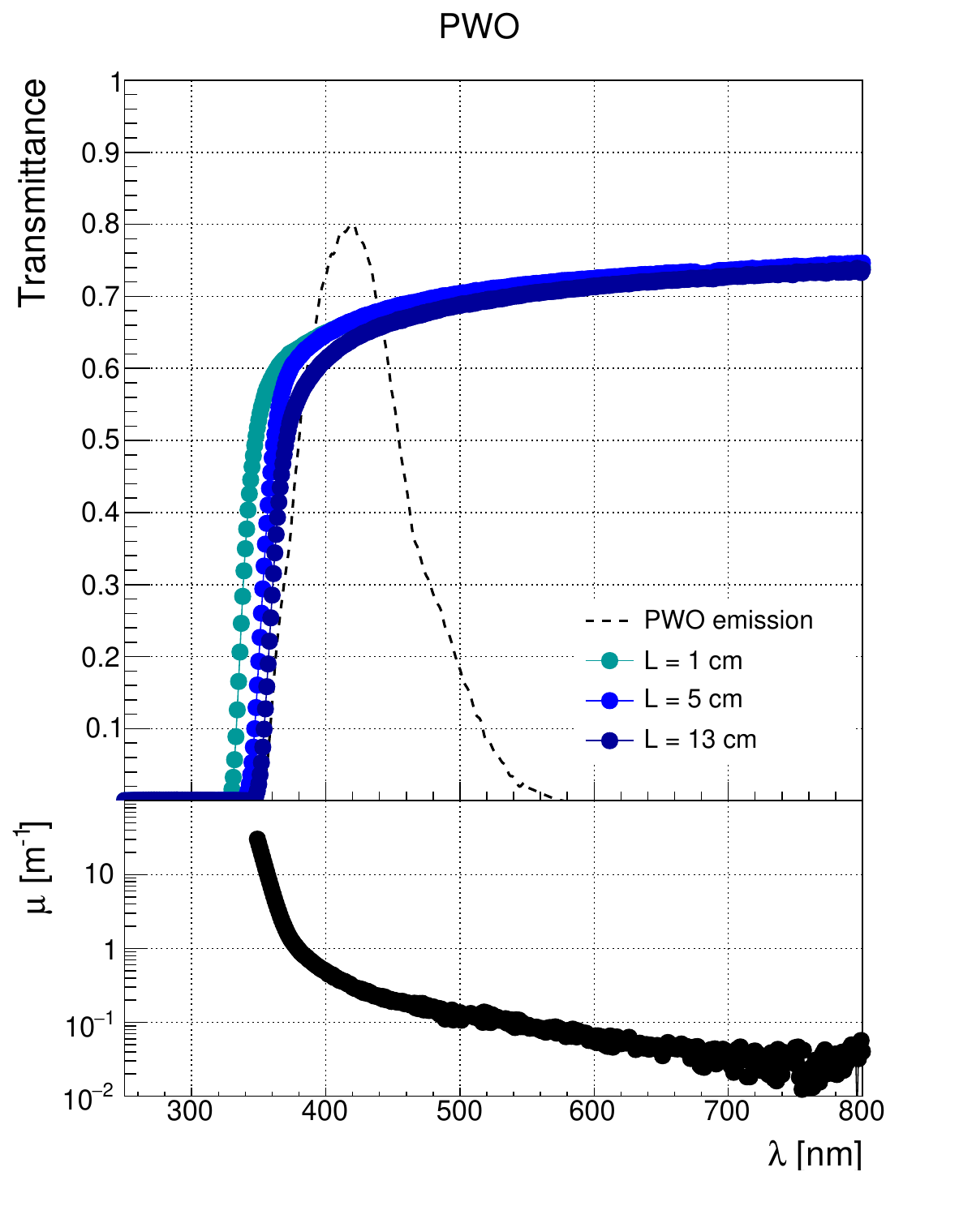}
    \caption{Transmittance curves are shown for BGO, BSO and PWO crystals of different length and their respective absorption coefficients are calculated using equation \ref{eq:mu_intr}. The emission spectra of BGO, BSO and PWO are also reported with dashed lines respectively on the left, central and right panel.}
    \label{fig:transmittance}
\end{figure}

The transmittance of each crystal was measured using a focused light beam passing through the sample along its longitudinal axis provided by a Perkin Elmer Lambda 650 UV/VIS spectrophotometer. While Fresnel losses at the surface are the same for each crystal length, the comparison of the transmittance of crystals with different lengths can provide an estimate of the intrinsic absorption coefficient as:
\begin{equation}\label{eq:mu_intr}
    \mu(\lambda) = \frac{1}{L_{1}-L_{2}} \ln\left(\frac{T_1(\lambda)}{T_2(\lambda)}\right)
\end{equation}
where we used $L_1 = 0.13~\rm m$, $L_2 = 0.01~\rm m$ and $T_1$ and $T_2$ represent the transmittance of the corresponding crystal at a certain wavelength, $\lambda$.
As shown in Figure~\ref{fig:transmittance}, for PWO and BGO crystals, we observe a comparable absorption coefficient at the emission peak (420~nm and 450~nm, respectively) of about 0.3~m$^{-1}$. We noticed instead a much larger absorption in BSO crystals (about $8~\rm m^{-1}$ at 450~nm) which %, after discussion with the manufacturer, 
has been attributed to the presence of impurities in the raw material used for the crystal growth and suggests the need of additional R\&D on the production process of this scintillator.

\subsection{Scintillation time profile}
\begin{figure}[!tbp]
    \centering
    \includegraphics[width=0.325\linewidth]{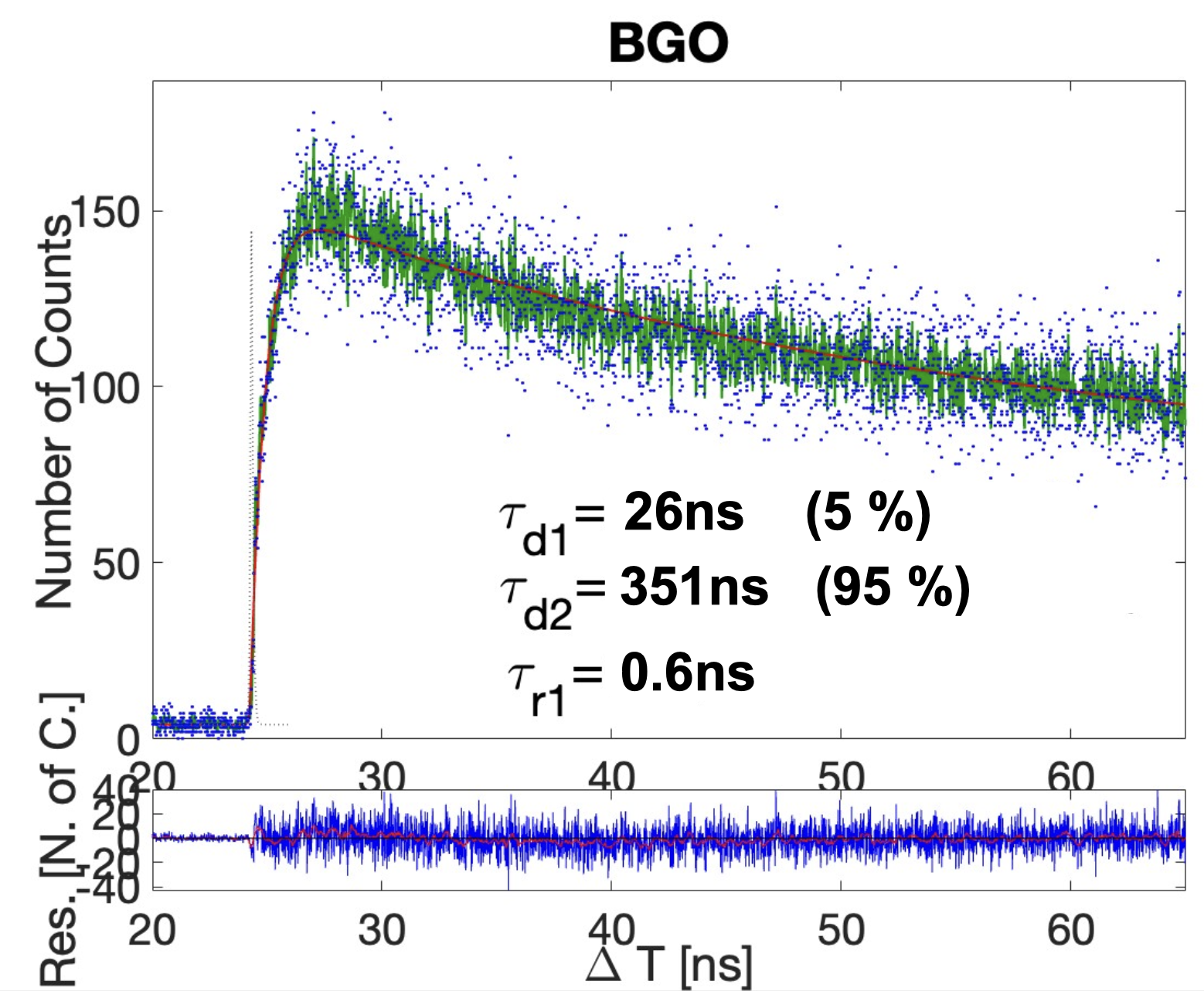}
    \includegraphics[width=0.325\linewidth]{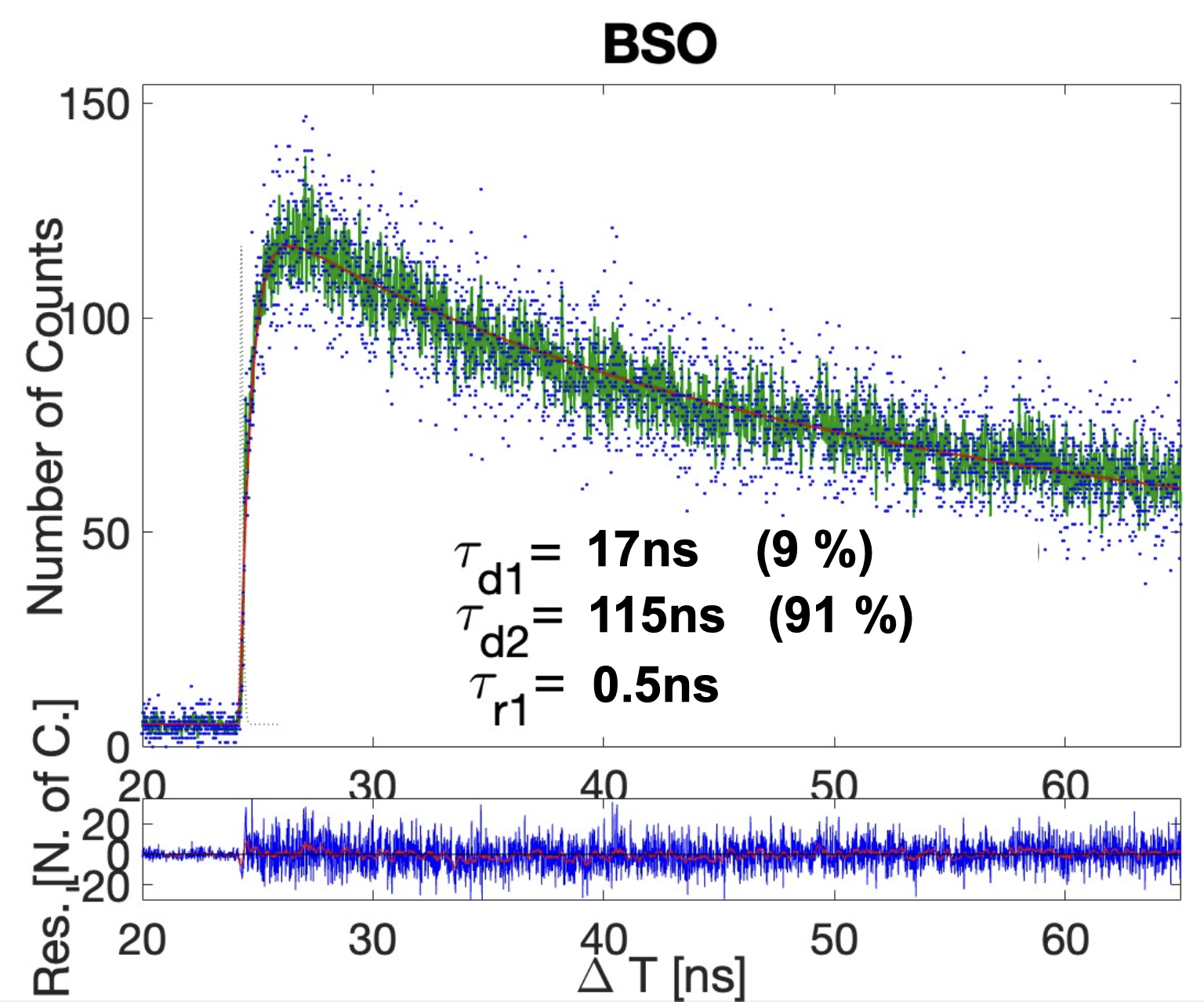}
    \includegraphics[width=0.325\linewidth]{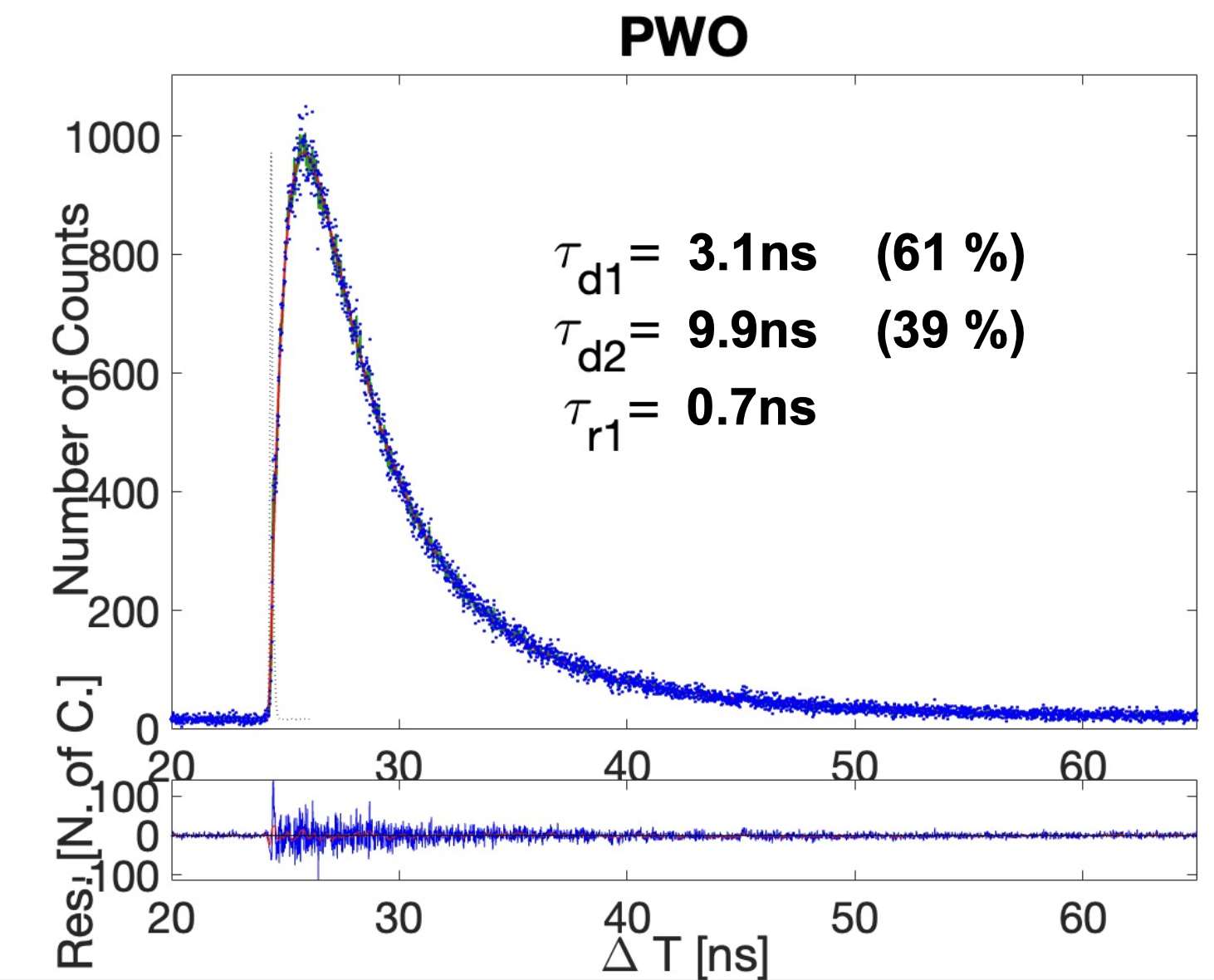}
    \caption{Scintillation decay time profile of the $1$~cm$^3$ BGO, BSO and PWO crystals (from left to right). The red curves represent the result of a fit using Eq.~\ref{eq:eff_decay_time}. The residuals of the distribution with respect to the red curve are reported in the bottom panels.}
    \label{fig:decay_time}
\end{figure}

The scintillation emission kinetics of the $1$~cm$^3$ BGO, BSO and PWO samples have been measured using a Time Correlated Single Photon Counting (TCSPC) method \cite{pagano_2022_decay}. The crystals were excited using X-rays generated from the excitation of an X-ray tube with a pulsed laser diode for 2 hours, which also provided the start signal. A hybrid photomultiplier (Becker \& Hickl HPM 100-07) located at one end of the crystal was used to count the time of arrival of single scintillation photons and to provide the stop signal.
The measured time profiles of the scintillation were fitted using equation \ref{eq:scinti_fit}, that includes one exponential rise time component, $\tau_r$, and two exponential decay components, $\tau_1, \tau_2$ with relative intensities $\rho_1, \rho_2$, as shown in Figure~\ref{fig:decay_time}.
\begin{equation}\label{eq:scinti_fit}
f(t\mid\theta)=\sum_{i=1}^{2}\,\rho_{i}\cdot\frac{\exp\Bigl(-\frac{t-\theta}{\tau_{d,i}}\Bigr)-\exp\bigl(-\frac{t-\theta}{\tau_{r}}\bigr)}{\tau_{d,i}-\tau_{r}}\cdot\Theta(t-\theta)+ \varepsilon
\end{equation}

The results of the fit are reported in Table~\ref{tab:decay_time}.
All crystals feature a rise time in the range 0.5-0.7~ns, most likely dominated by the transit time of photons inside the crystal. An effective decay time (defined in equation \ref{eq:eff_decay_time}) of 217, 76 and 4~ns was measured for the BGO, BSO and PWO samples, respectively.

\begin{equation}\label{eq:eff_decay_time}
    \tau_{\mathrm{d, eff}}^{-1} = \sum_{i=1}^{n=2} \frac{\rho_i}{\tau_{\mathrm{d,i}}} 
\end{equation}

\begin{table}[!tbp]
\centering \caption{Scintillation kinetics parameters measured for the $1$~cm$^3$ BGO, BSO and PWO crystals.} \vspace{0.2cm}
\begin{tabular}{c|c|c|c|c|c|c}
\hline
Crystal & $\tau_r$ [ns]  & $\tau_{d,1}$  [ns]   & $\rho_1$  [\%]   & $\tau_{d,2}$  [ns]   & $\rho_2$   [\%]  & $\tau_{\mathrm{d,eff}}$  [ns]\\ 
\hline \hline
BGO     &  0.6$\pm0.1$      & 26$\pm2$           & 5$\pm0.5$          &  351$\pm4$             & 95$\pm1$          &  217$\pm4$    \\
\hline
BSO     &  0.5$\pm0.1$      & 17$\pm2$            & 9$\pm1$          &  115$\pm2$             &  91$\pm1$         &  76$\pm1$    \\ 
PWO     &  0.7$\pm0.1$      & 3.1$\pm0.3$            & 61$\pm1$         &   9.9$\pm0.6$            &   39$\pm2$          &  4$\pm0.3$    \\ 
\hline
\end{tabular} 
\label{tab:decay_time}
\end{table}
\vspace{0.6cm}

\subsection{Scintillation light output}
\begin{figure}[!tbp]
    \centering
    \includegraphics[width=0.495\linewidth]{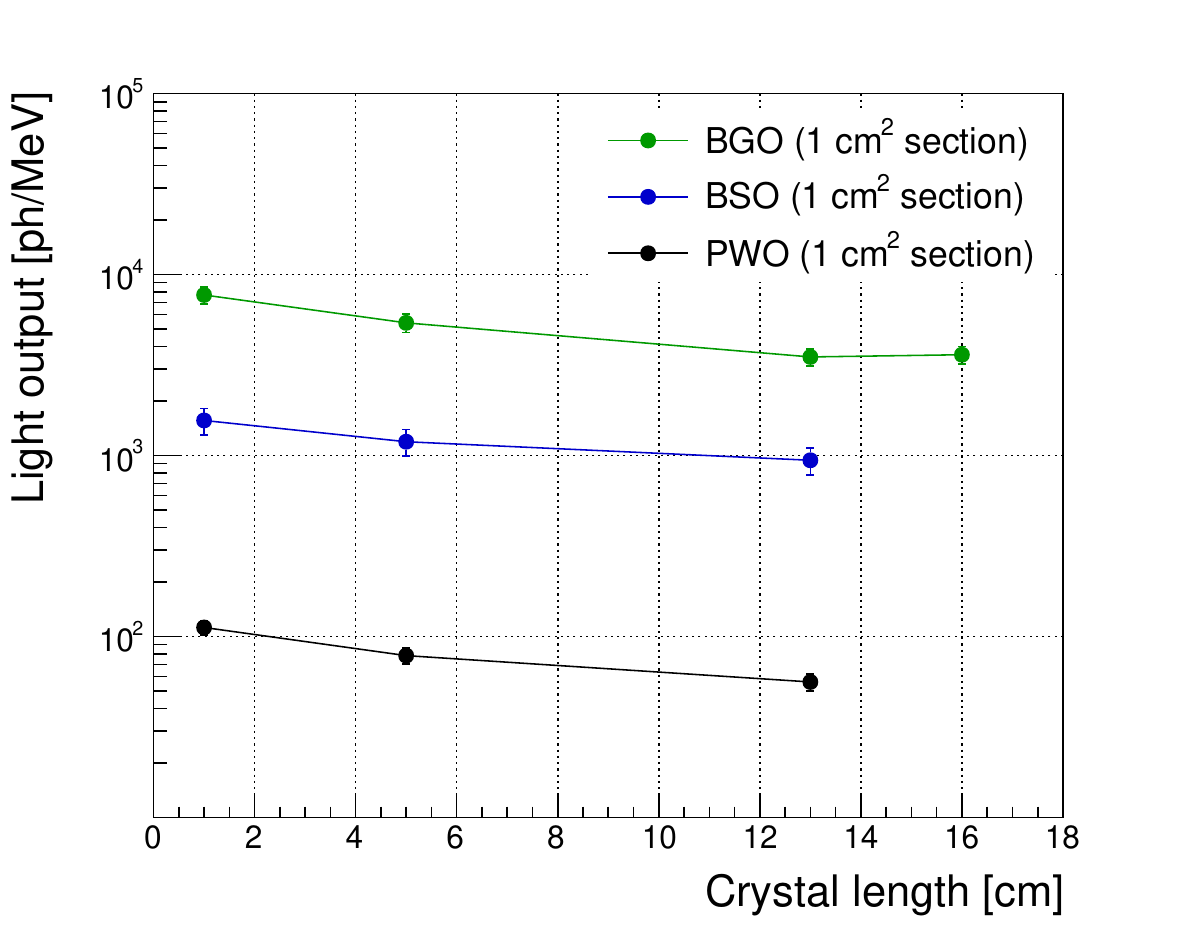}
    \includegraphics[width=0.495\linewidth]{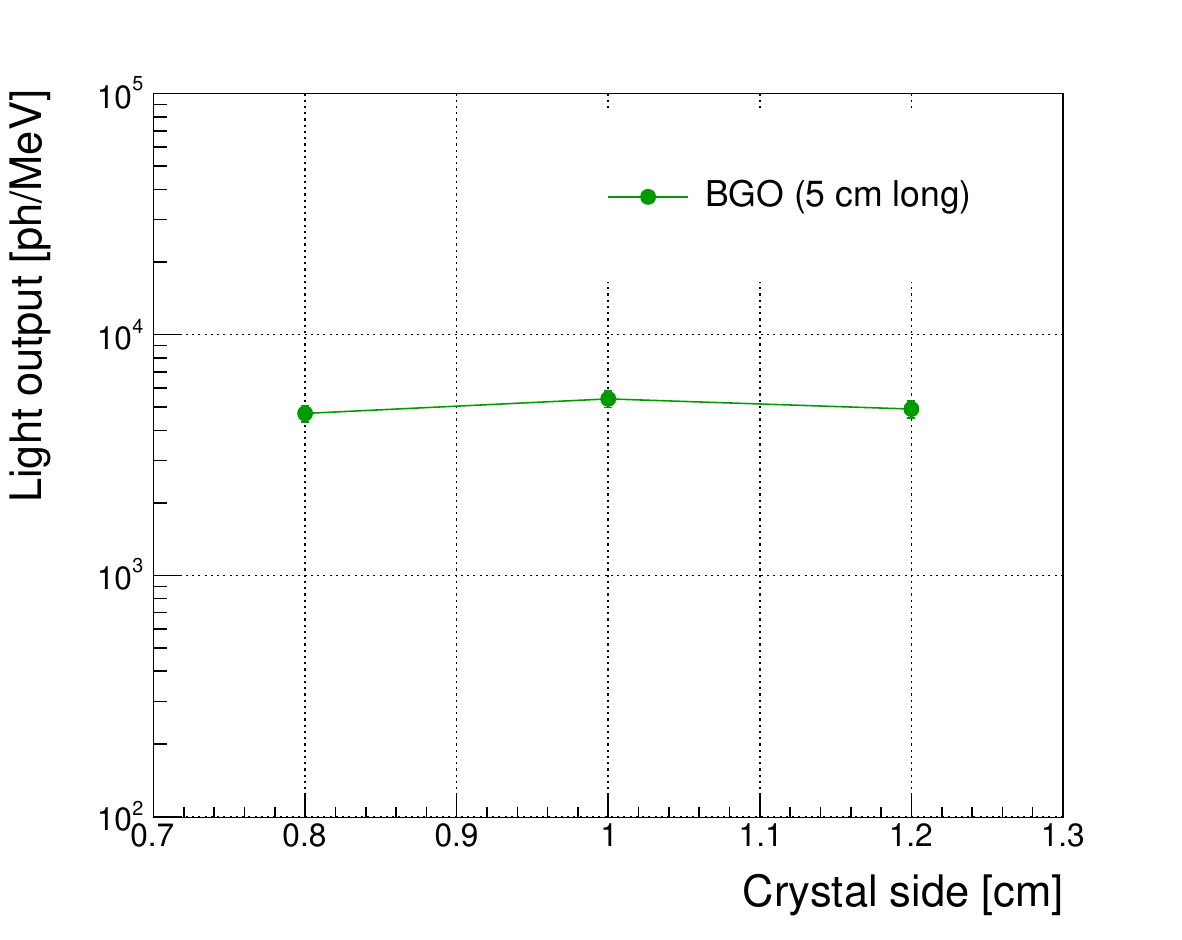}
    \caption{Left: Light output of PWO, BGO and BSO crystals with $1\times 1~\rm cm^2$ section and different lengths measured with a PMT (Hamamatsu R2059). Right: light output of 5~cm long BGO crystals with different sections.}
    \label{fig:lo_pmt}
\end{figure}

The scintillation light output of each crystal was measured by coupling one end of the crystal to a photomultiplier tube (Hamamatsu R2059 PMT) with a circular photocathode of approximately 3 cm in diameter, using optical grease (Rhodorsil 55), ensuring that the entire crystal face was read out. All the other crystal surfaces were fully wrapped with several Teflon layers. The PMT signal was measured with a CAEN DT5720 digitizer, integrated over a 4~$\mu s$ gate and converted in number of photoelectrons using the estimation of the single photo electron response. The PMT quantum efficiency, weighted over the emission spectrum of the corresponding crystal, was then used to calculate the light output in terms of photons per unit of energy (ph/MeV).
As shown in Figure~\ref{fig:lo_pmt}, all crystals feature a decrease of the light output of about 50\% from 10 to 130~mm length. On the other hand, varying the crystal cross section has no impact on the light output as long as the PMT active area is always larger than the crystal face so that all light coming out of the crystal is detected. For 130~mm long BGO, BSO and PWO crystals with $1\times 1~\rm cm^2$ section, a light output of $(3500\pm390)$, $(950\pm110)$ and $(60\pm10)$ photons/MeV, respectively, was measured. The error on the absolute light output was estimated as the combination of a 5\% uncertainty on the single photoelectron calibration and a 10\% uncertainty on the PMT quantum efficiency.

\begin{figure}[!tbp]
    \centering
    \includegraphics[width=0.495\linewidth]{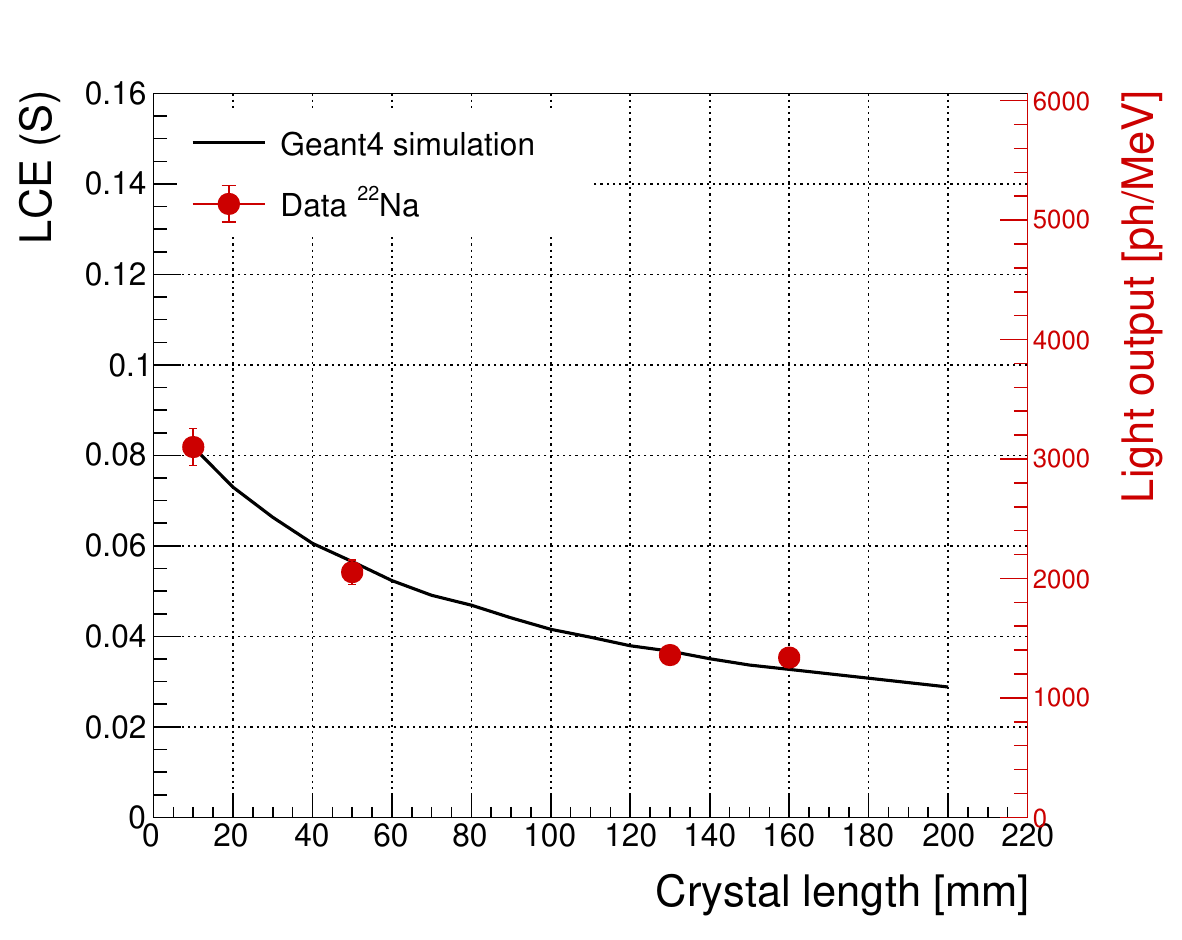}
    \includegraphics[width=0.495\linewidth]{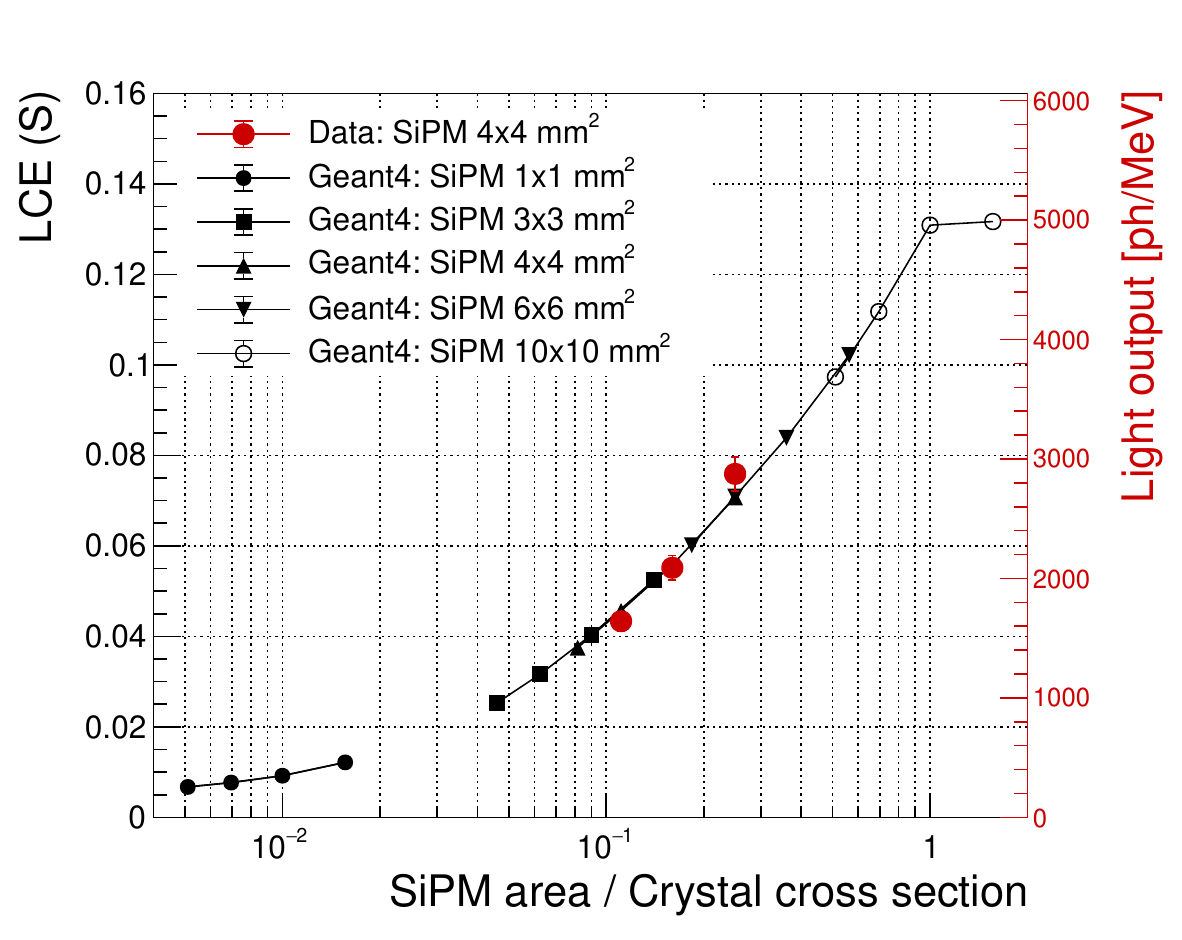}
    \caption{Light output of BGO crystals readout with a $4\times 4$~mm$^2$ SiPM measured with a $^{22}$Na source and compared with prediction from Geant4 ray-tracing simulation. Results are shown for $10\times10$~mm$^2$ section crystals of different lengths (left panel) and for 5~cm long crystals of different sections (right panel).}
    \label{fig:lce_w_sipms}
\end{figure}

In the calorimeter concept proposed in \cite{Lucchini_2020}, the crystal readout is performed using SiPMs with an active area smaller than the crystal face. Therefore, it is crucial to assess how the light output of a given crystal scales with the fraction of its surface covered by the photodetector.
We performed this study using the three 5~cm long BGO samples with different sections ($8\times 8$~mm$^2$, $10\times 10$~mm$^2$ and $12\times 12$~mm$^2$) and different lengths and by reading them out with a silicon photomultiplier from Advansid (NUV-HD, 40$\rm~\mu m$ cell size) with an active area of $4\times 4$~mm$^2$.
As shown in Figure~\ref{fig:lce_w_sipms}, the results are in good agreement with the Geant4 ray-tracing simulation of the setup and provide a quantitative method to determine the minimum SiPM size required for each crystal to ensure a scintillation signal above 2000~phe/GeV.

\section{Optical filter transmittance}\label{sec:filters_trans}
A set of optical filters designed to separate Cherenkov and scintillation light by absorbing photons in the wavelength region around the scintillation emission peak was tested.
We compared the performance of two types of optical filters -- interference and absorptive filters -- as summarized in Table \ref{tab:filter_list}.
Interference (or dichroic) filters are intrinsically thin because they rely on thin-film interference to transmit a selected wavelength band while reflecting or suppressing other wavelengths. However, their transmittance curve inevitably depends on the angle of incidence, since the phase difference between the incident and reflected waves varies with angle \cite{Bacon_2025}.
The latter filters instead typically consist of colored glass or pigmented gelatin resin and absorb optical photons in a specific range of wavelength with a nearly angular-independent behavior. The probability of absorbing photons depends on the amount of pigmentation or dye on the filter and also on the filter thickness. Furthermore, due to the materials used to manufacture these filters, they can produce fluorescence photons \cite{filter_fluorescence}.

While a broad variety of filters is available off-the-shelf we have focused on filters with thickness below 0.3~mm which are best suited for integration in a full scale calorimeter concept because of two main reasons: they could potentially be integrated within (or in place of) the SiPM protective window and they provide a better light collection efficiency compared to thicker filters ($>1$~mm) for which more Cherenkov photons would escape from the sides, in the gap between the crystal and the SiPM.

The two interference filters tested (Eve-Int-580, Eve-Int-650), were custom designed by Everix to cut off the wavelengths corresponding to the PWO and BGO/BSO scintillation emission, respectively in the range of 380--580~nm and 380--660~nm at normal incidence.
The six thin long-pass absorptive filters tested (5 from Kodak and 1 from Everix) were instead selected with a cut-off wavelength optimal for PWO crystals. 
For comparison, we also tested two more conventional and thicker absorptive glass filters produced from Hoya: O56 and U330.

\begin{figure}[!tbp]
    \centering
    \includegraphics[height=0.16\linewidth]{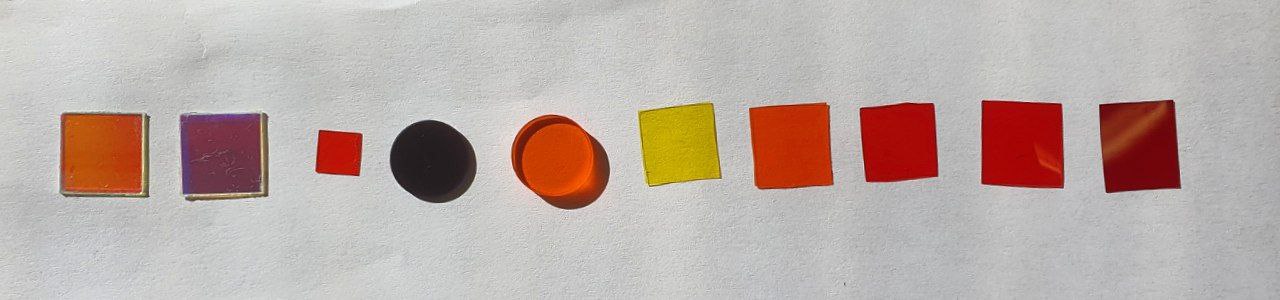}
    \includegraphics[height=0.16\linewidth]{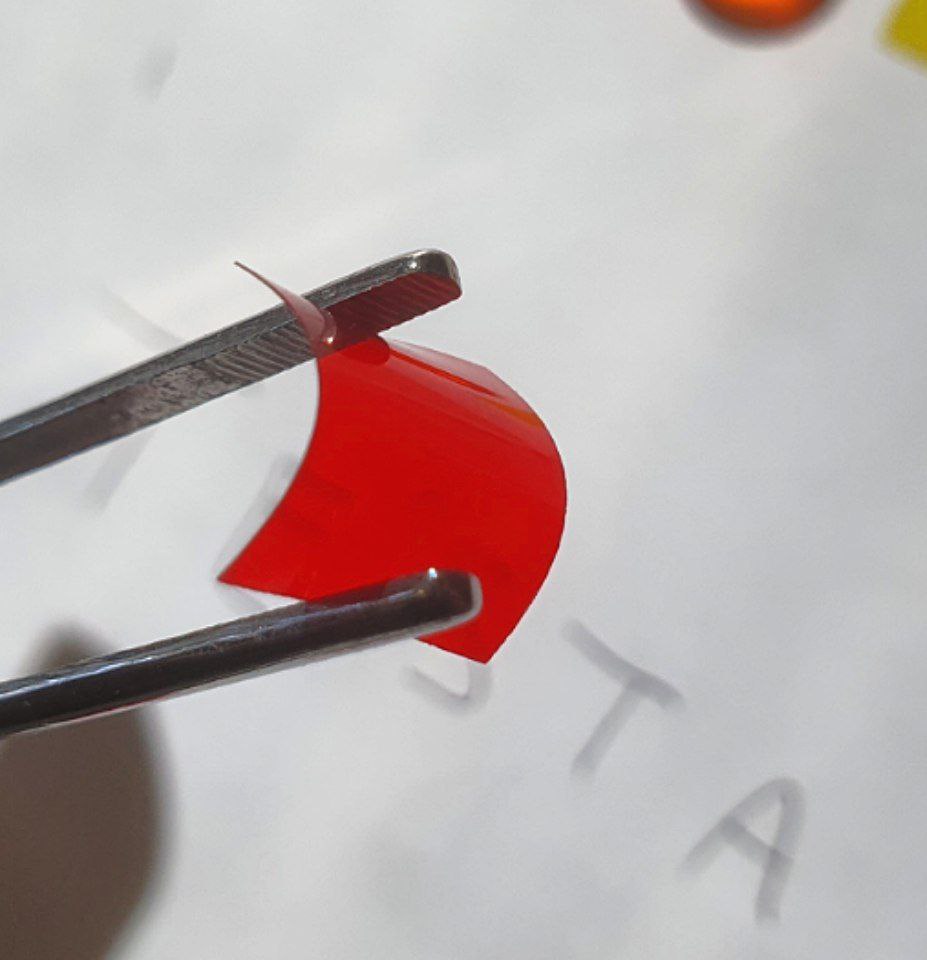}
    \includegraphics[height=0.16\linewidth]{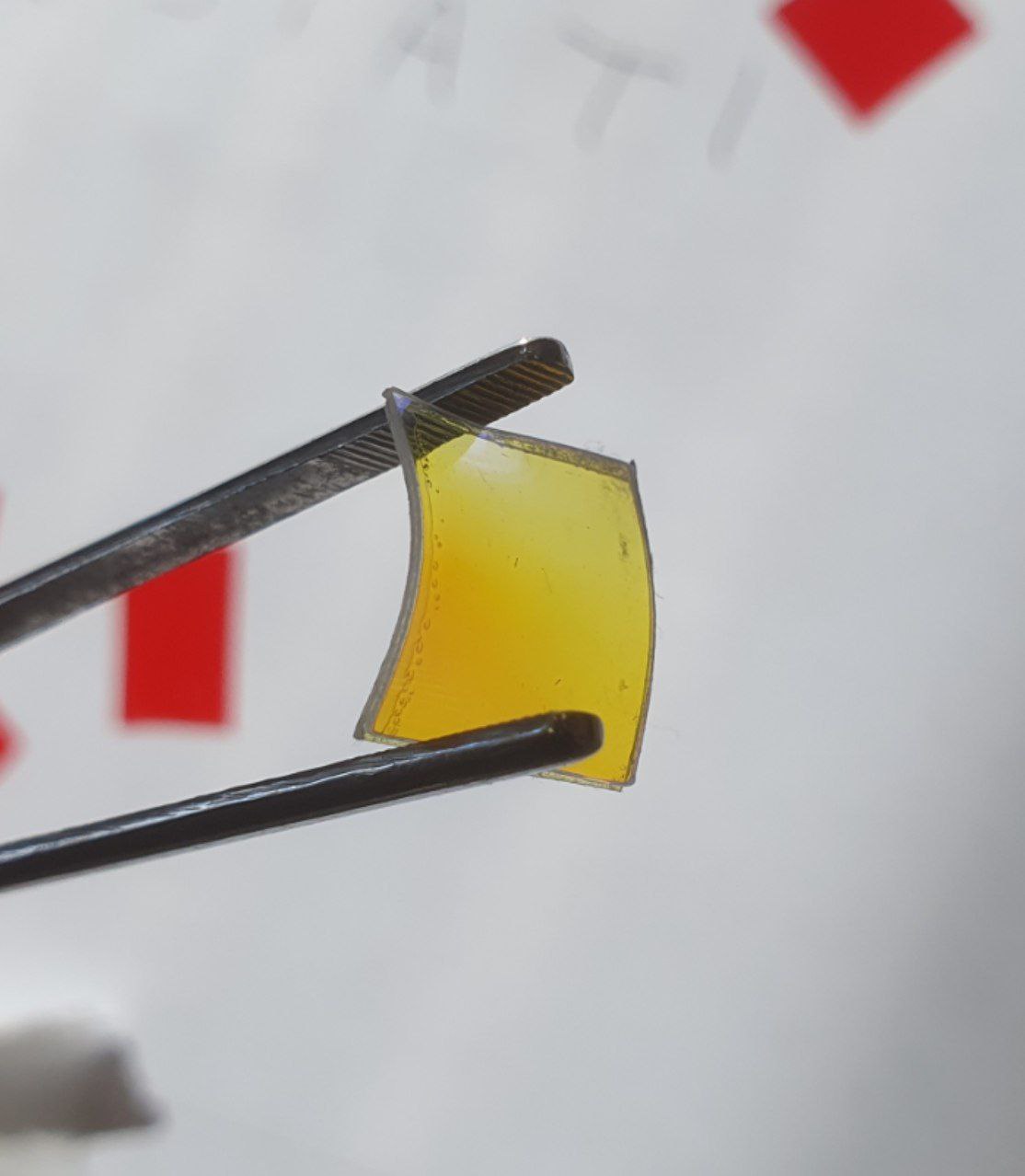}
    \caption{Picture of the filters tested. From left to right: Everix-Int-580, Everix-Int-650, Eve-Abs-580, Hoya-U330, Hoya-O56, Kodak-8, Kodak-22, Kodak-24, Kodak-25, Kodak-29. Side views of the Kodak-24 (absorptive) and the Everix-Int-580 (interference) filter being warped.}
    \label{fig:pictures_filters}
\end{figure}

\begin{table}[!tbp]
\centering \caption{List of filters tested.} \vspace{0.2cm}
\begin{tabular}{l|c|c|c|c}
\hline
Filter label &    Manufacturer  & Thickness    & Cut-off range  & Type  \\ 
\hline\hline
Hoya-U330    &    Hoya          & 1.00 mm       & $400<\lambda<680$~nm          & Absorptive  \\ 
Hoya-O56     &    Hoya          & 2.50 mm       & $\lambda<560$~nm   & Absorptive  \\ 
\hline
Kodak-8      &    Kodak         & 0.10 mm       & $\lambda<490$~nm   & Absorptive  \\ 
Kodak-22     &    Kodak         & 0.10 mm       & $\lambda<565$~nm   & Absorptive  \\ 
Kodak-24     &    Kodak         & 0.10 mm       & $\lambda<590$~nm   & Absorptive  \\ 
Kodak-25     &    Kodak         & 0.10 mm       & $\lambda<595$~nm   & Absorptive  \\ 
Kodak-29     &    Kodak         & 0.10 mm       & $\lambda<620$~nm   & Absorptive  \\ 
\hline
Eve-Abs-580  &    Everix        & 0.26 mm       & $\lambda<580$~nm   & Absorptive  \\ 
Eve-Int-580  &    Everix        & 0.26 mm       & $370<\lambda<580$~nm   & Interference \\ 
Eve-Int-650  &    Everix        & 0.26 mm       & $370<\lambda<650$~nm   & Interference \\ 
\hline
\end{tabular} 
\label{tab:filter_list}
\end{table}
\vspace{0.6cm}

\begin{figure}[!tbp]
    \centering
    \includegraphics[width=0.495\linewidth]{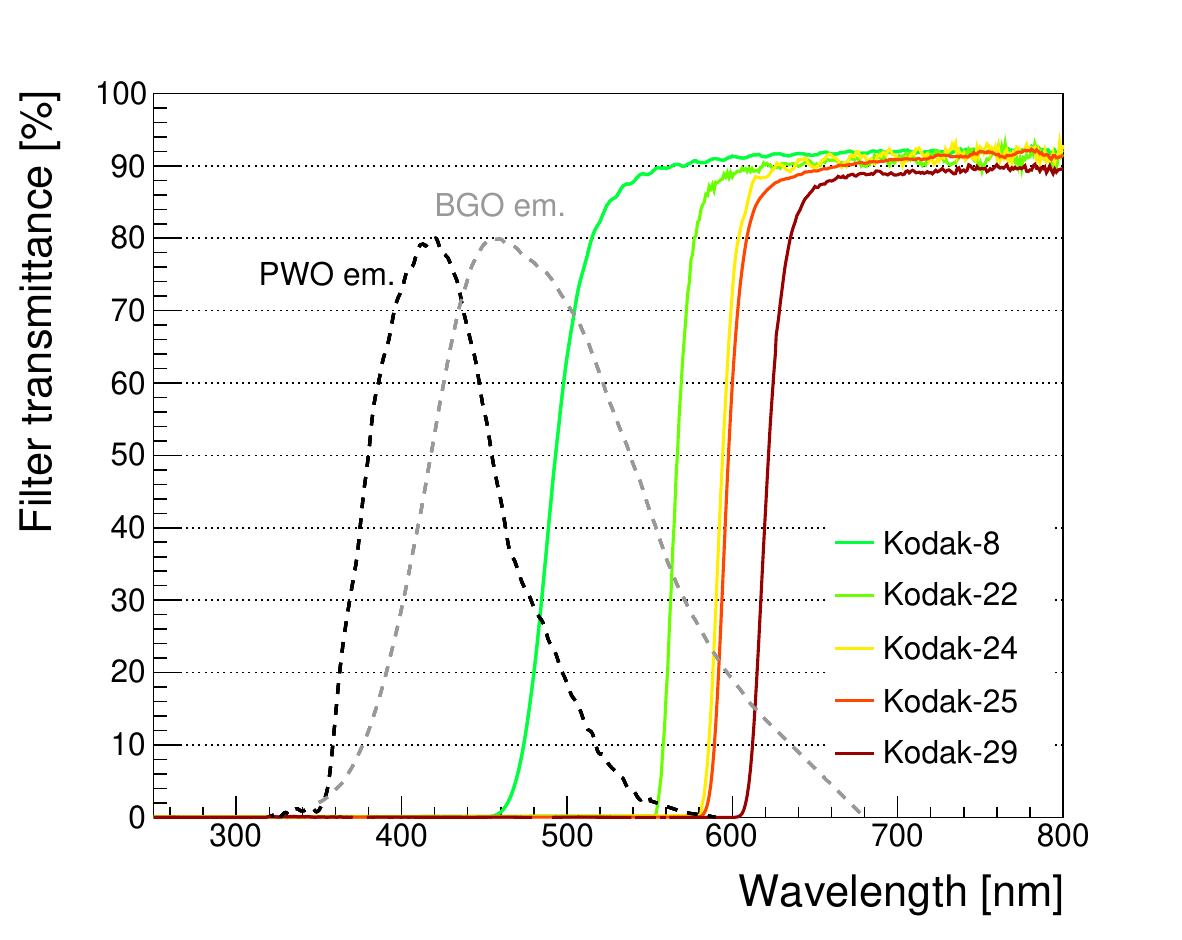}
    \includegraphics[width=0.495\linewidth]{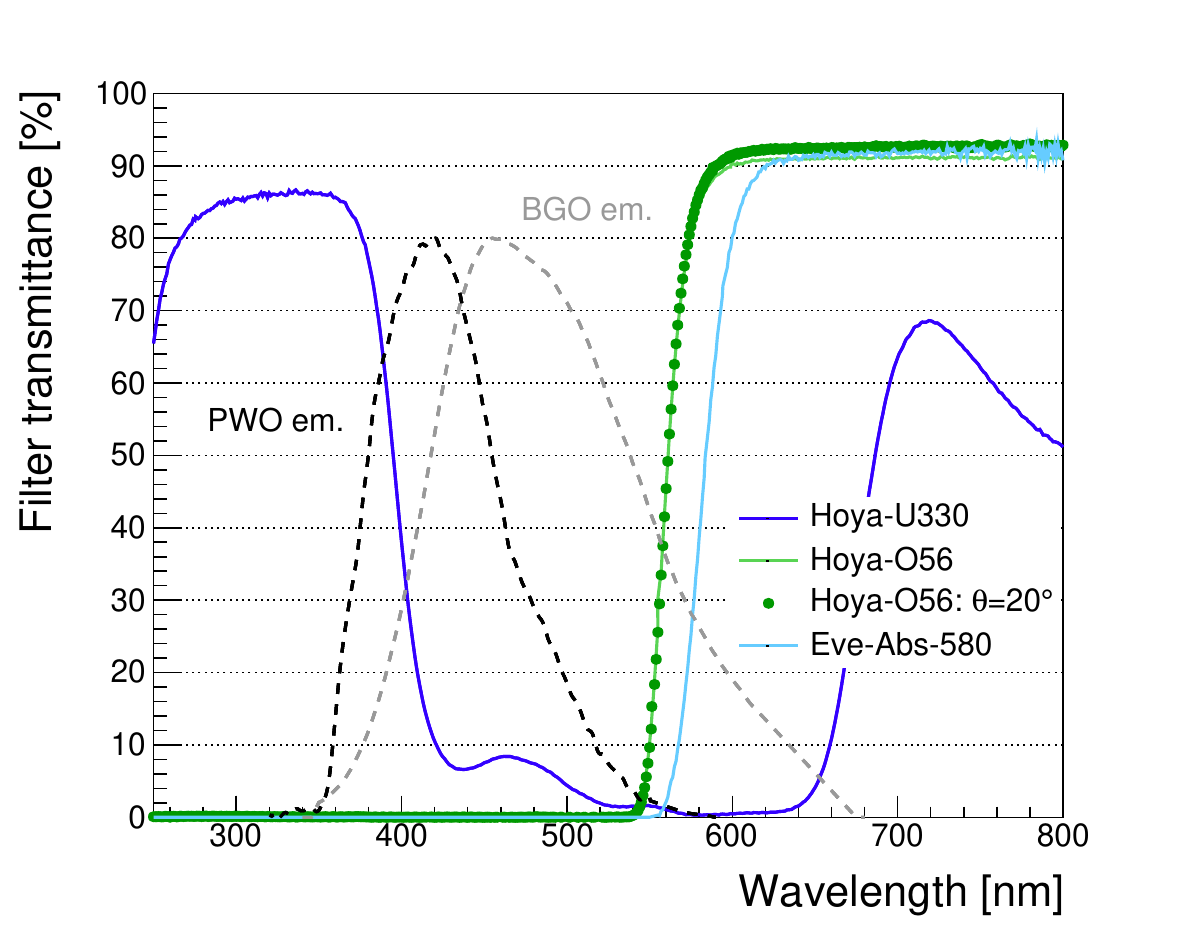}
    \caption{Left: transmittance curves of the Kodak absorptive filters at normal incidence. Right: transmittance curves of the Hoya and Everix absorptive filters at normal incidence and of the Hoya-O56 filter measured at 20$^{\circ}$ incidence angle (green dots). PWO and BGO emission spectra are superimposed for comparison.}
    \label{fig:absorptive_filters}
\end{figure}

\begin{figure}[!tbp]
    \centering
    \includegraphics[width=0.495\linewidth]{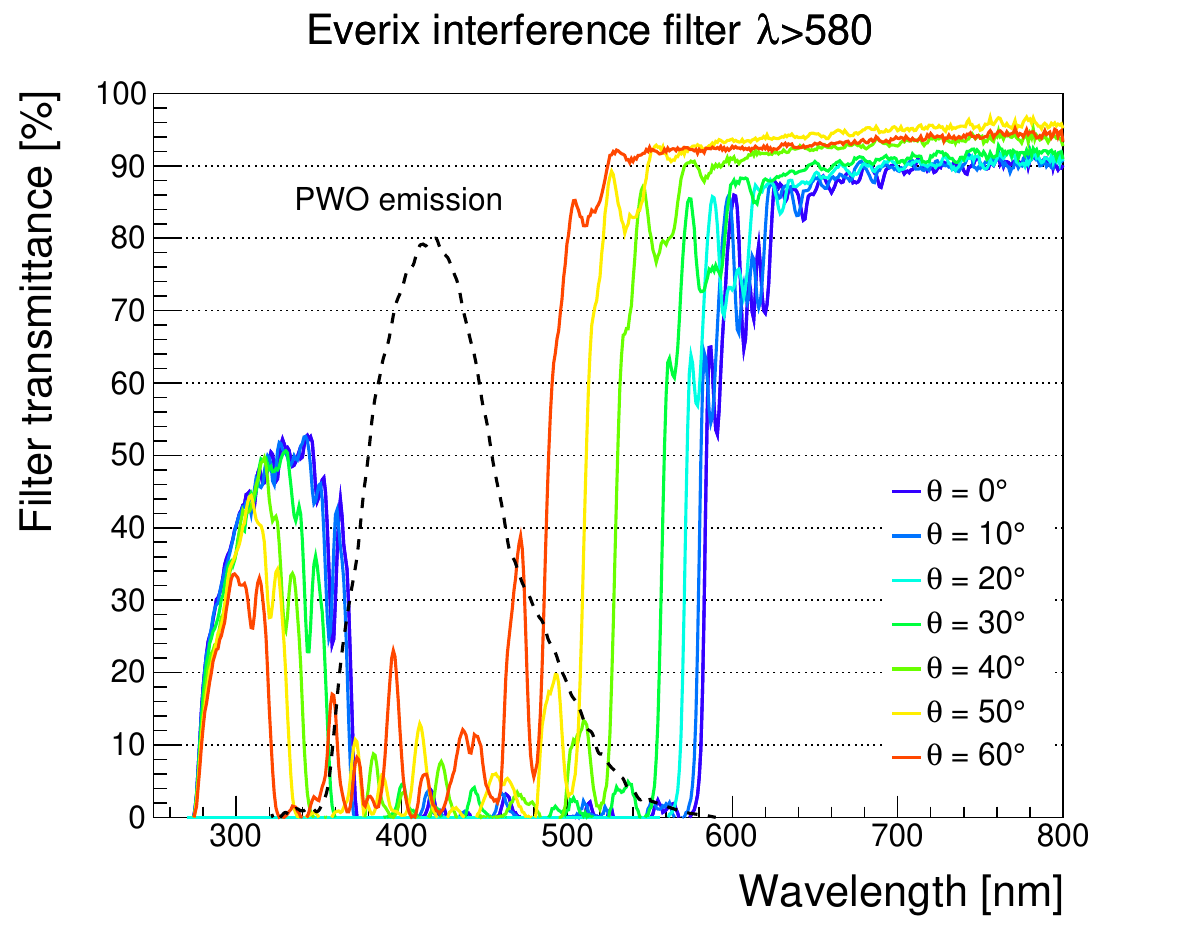}
    \includegraphics[width=0.495\linewidth]{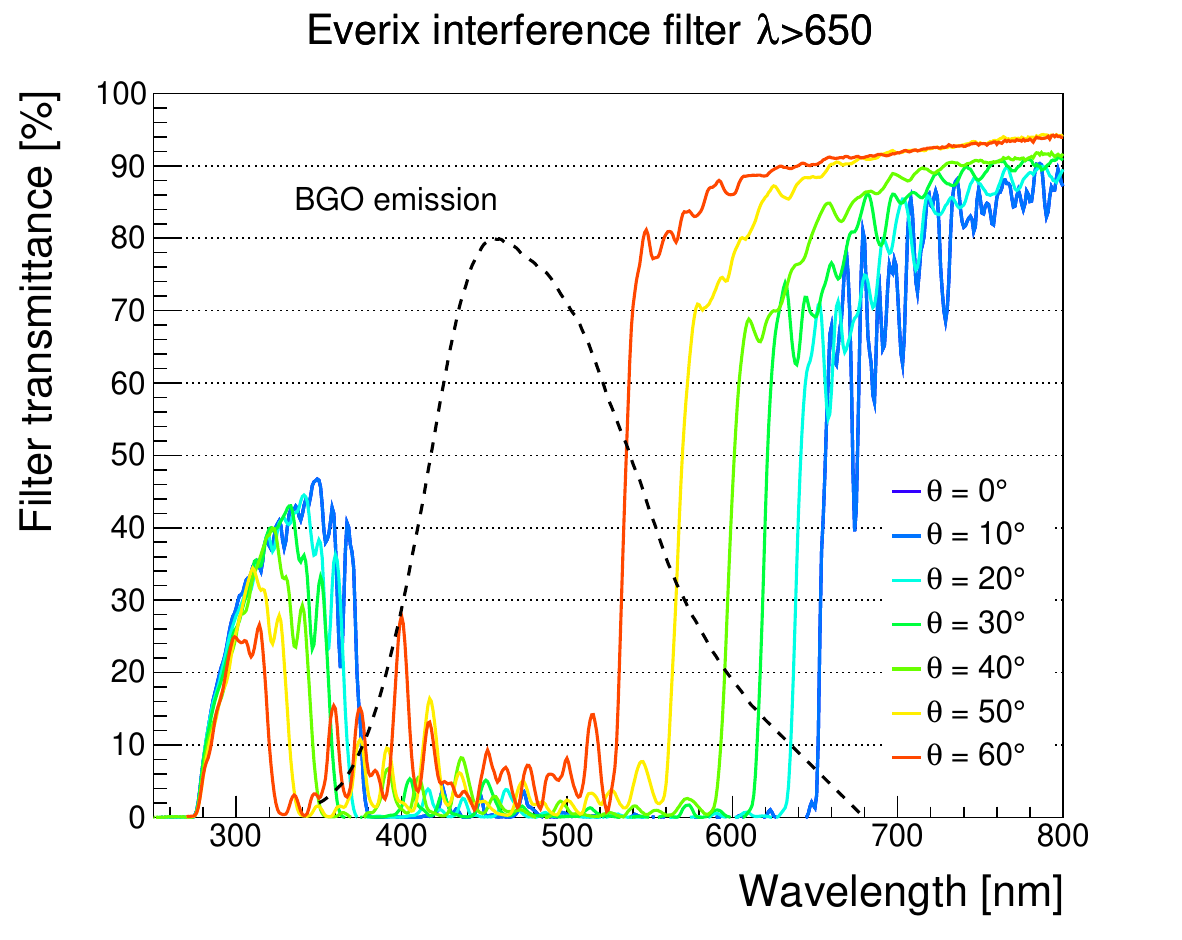}
    \caption{Left: transmittance curves of the 580~nm Everix interference filter (Eve-Int-580) for different angles of incident light beam, compared with PWO emission spectrum. Right: transmittance curves of the 650~nm Everix interference filter (Eve-Int-650) for different angles of incident light beam, compared with BGO emission spectrum.}
    \label{fig:interference_filters}
\end{figure}

The filter transmittance was measured with a spectrophotometer (Perkin Elmer Lambda 650 UV/VIS) for different incidence angles of the optical beam. The results are reported in Figure~\ref{fig:absorptive_filters} for the Everix, Hoya and Kodak absorptive filters measured at normal incidence. The Hoya-U330 and Kodak-29 filters appear as more suitable to filter out the scintillation from BGO and BSO crystals while the other filters are better matched to the PWO emission.
For all these filters, we did not observe any change in the cut-off wavelength when changing the impact angle of photons. We report for illustration an example of the Hoya-O56 filter measured at 20$^{\circ}$ incidence angle in Figure~\ref{fig:absorptive_filters}.
In absorptive filters photons are absorbed by the material (and in some cases re-emitted at a random angle) and the probability of a photon to be absorbed does not show a significant dependence on the photon angle. %(say that overall light attenuation is larger and larger angles because of a longer path inside the material but this affects mostly the optical density rather than the cut-off wavelength.)
Conversely, and as expected, we observed a strong angular dependence of the transmittance curves for interference filters as shown in Figure~\ref{fig:interference_filters}. This effect is intrinsic to the functional principle of thin-film interference, in which light is reflected by layers with different refractive indices, giving rise to destructive interference depending on the phase difference between the incident and reflected waves. When the angle of incidence changes, the effective optical path length between two layers varies, which in turn modifies the phase difference and leads to a shift in the wavelength at which destructive interference occurs. A recent detailed optical model was presented in \cite{Bacon_2025}.

\section{Filter performance in shielding scintillation light}\label{sec:filter_validation}

Based on the emission spectra of scintillation light, $\rm EM(\lambda)$ and crystal transmittance, $T_c(\lambda)$, presented in Section~\ref{sec:crystal_char} and the transmission curves of optical filters, $T_f(\lambda)$, discussed in Section~\ref{sec:filters_trans}, it is possible to estimate the fraction of scintillation light that is expected to pass through the filter as:
\begin{equation}\label{eq:scint_ratio}
    \rm \frac{S}{S_0} = \frac{\int_{300~nm}^{800~nm} EM(\lambda)\cdot T_c(\lambda)\cdot T_f(\lambda) d\lambda }{\int_{300~nm}^{800~nm} EM(\lambda) \cdot T_c(\lambda) d\lambda }
\end{equation}

\noindent
The expected fraction of light passing through a certain filter is shown in Table~\ref{tab:fraction_light}, calculated for BGO/BSO, PWO and LYSO crystals which feature an emission spectrum close to that of PWO, peaking at 425~nm \cite{Addesa_2022}.

\begin{table}[!tbp]
\centering \caption{Expected fraction of scintillation light ($S/S_0$) passing through a certain optical filter at normal incidence angle ($\theta = 0$).} \vspace{0.2cm}
\begin{tabular}{l|c|c|c}
\hline
Filter label &   ~ BGO/BSO~  & ~~~~~PWO~~~~~    & ~~~~~LYSO~~~~~  \\ 
\hline\hline
Hoya-U330 & 8.8\% & 26.8\% & 14.8\%\\
Hoya-O56 & 12.8\% & 0.3\% & 1.5\%\\
\hline
Kodak-8 & 40.7\% & 7.9\% & 9.2\%\\
Kodak-22 & 11.2\% & 0.1\% & 1.2\%\\
Kodak-24 & 6.2\% & <0.1\% & 0.6\%\\
Kodak-25 & 5.5\% & <0.1\% & 0.5\%\\
Kodak-29 & 2.7\% & <0.1\% & 0.3\%\\
\hline
Eve-Abs-580 & 8.1\% & <0.1\% & 0.8\%\\
Eve-Int-580 & 7.0\% & 1.4\% & 0.8\%\\
Eve-Int-650 & 1.1\% & 2.5\% & 0.7\%\\
\hline
\end{tabular} 
\label{tab:fraction_light}
\end{table}
\vspace{0.6cm}

To verify the validity of these predictions, which rely on the experimental measurements of the crystal scintillation emission spectrum and the filter transmittance, an experimental method based on a $^{22}$Na radioactive source was devised. A $3\times3\times5$~mm$^3$ LYSO:Ce crystal was used as a test sample because its light output is much higher than that of the other crystals, thus facilitating the measurement of the light yield with low-energy sources. Furthermore, its emission spectrum is very similar to that of PWO, making it a good proxy for these crystals in terms of filter performance.

\begin{figure}[!tbp]
    \centering
    \includegraphics[width=0.495\linewidth]{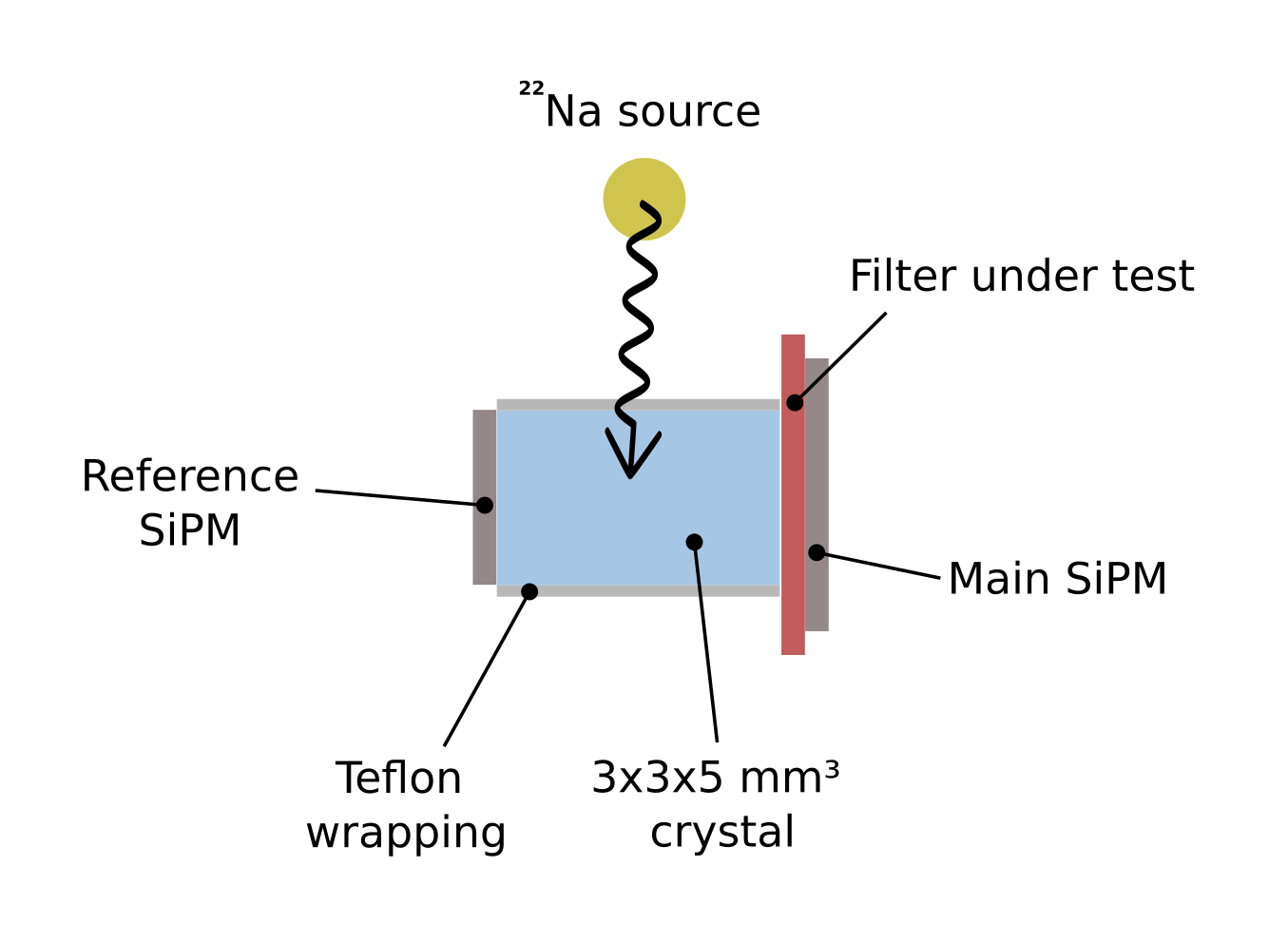}
    \includegraphics[width=0.495\linewidth]{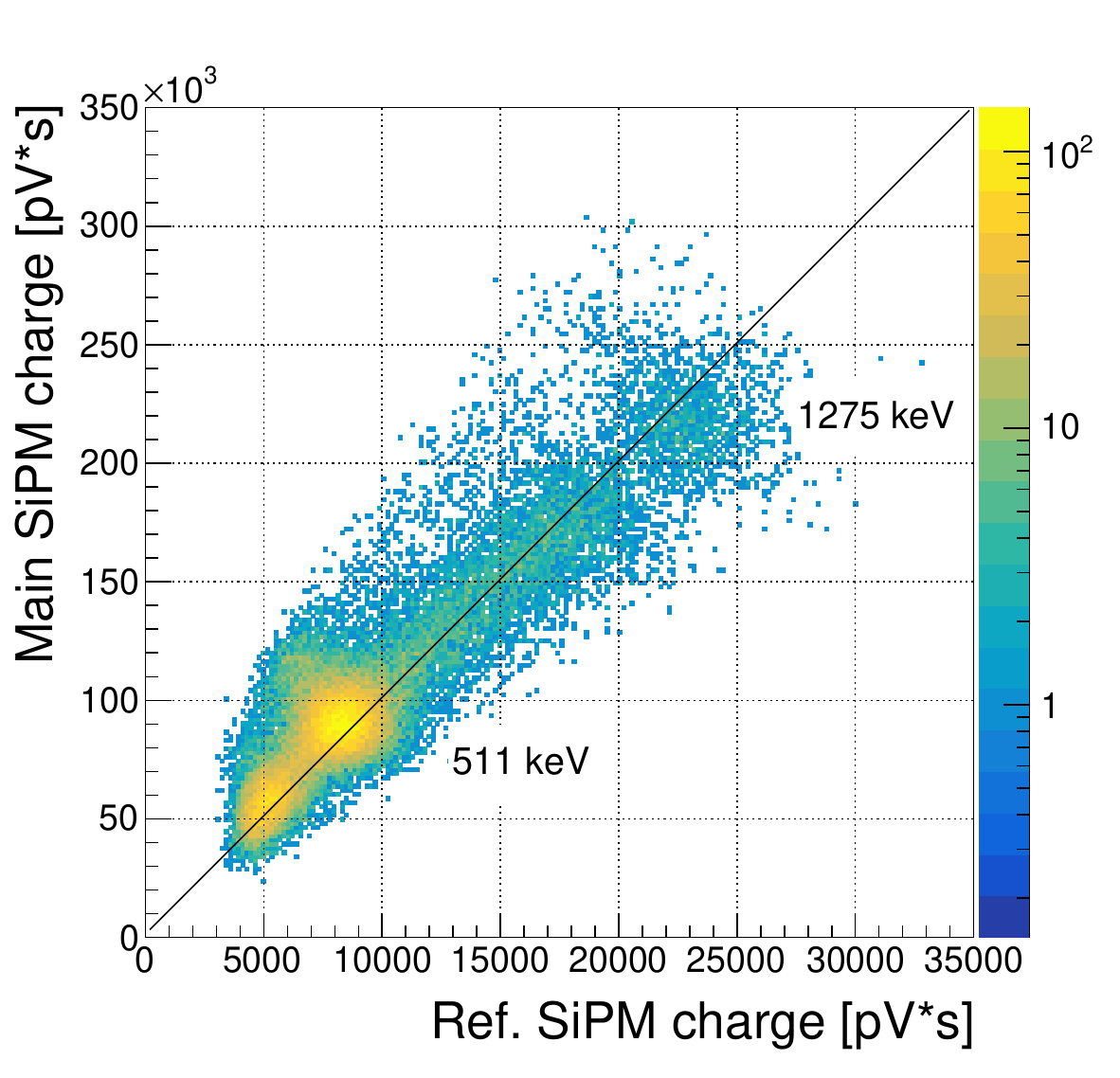}
    \caption{Left: schematics view of the experimental setup used for filter performance evaluation using radioactive $^{22}$Na source. Right: correlation of the $^{22}$Na source signal (integrated charge) between the reference and main SiPM.}
    \label{fig:na22setup}
\end{figure}

The LYSO sample was wrapped in Teflon and read out at the two opposite ends with two different SiPMs, as illustrated in Figure~\ref{fig:na22setup}. One SiPM (“reference”, model HPK S12572-015C) was glued to the crystal and used to tag gamma-ray interactions and estimate the deposited energy. The second SiPM (“main”, model HPK S13360-6050CS) was used to measure the amount of scintillation light detected without a filter and with different types of filters placed between the crystal and the SiPM. The main SiPM was operated at 53~V (2~V above the breakdown voltage), corresponding to a peak PDE of 32\% at 450~nm. No glue or other optical coupling was used for the main SiPM to minimize systematic uncertainties.

The SiPM signals were amplified using two transimpedance amplifiers mounted on Advansid evaluation boards, and the waveforms were recorded with an oscilloscope.
The right panel of Figure~\ref{fig:na22setup} shows the correlation between the integrated charge measured by the reference and main SiPMs without filter.

\begin{figure}[!tbp]
    \centering
    \includegraphics[width=0.325\linewidth]{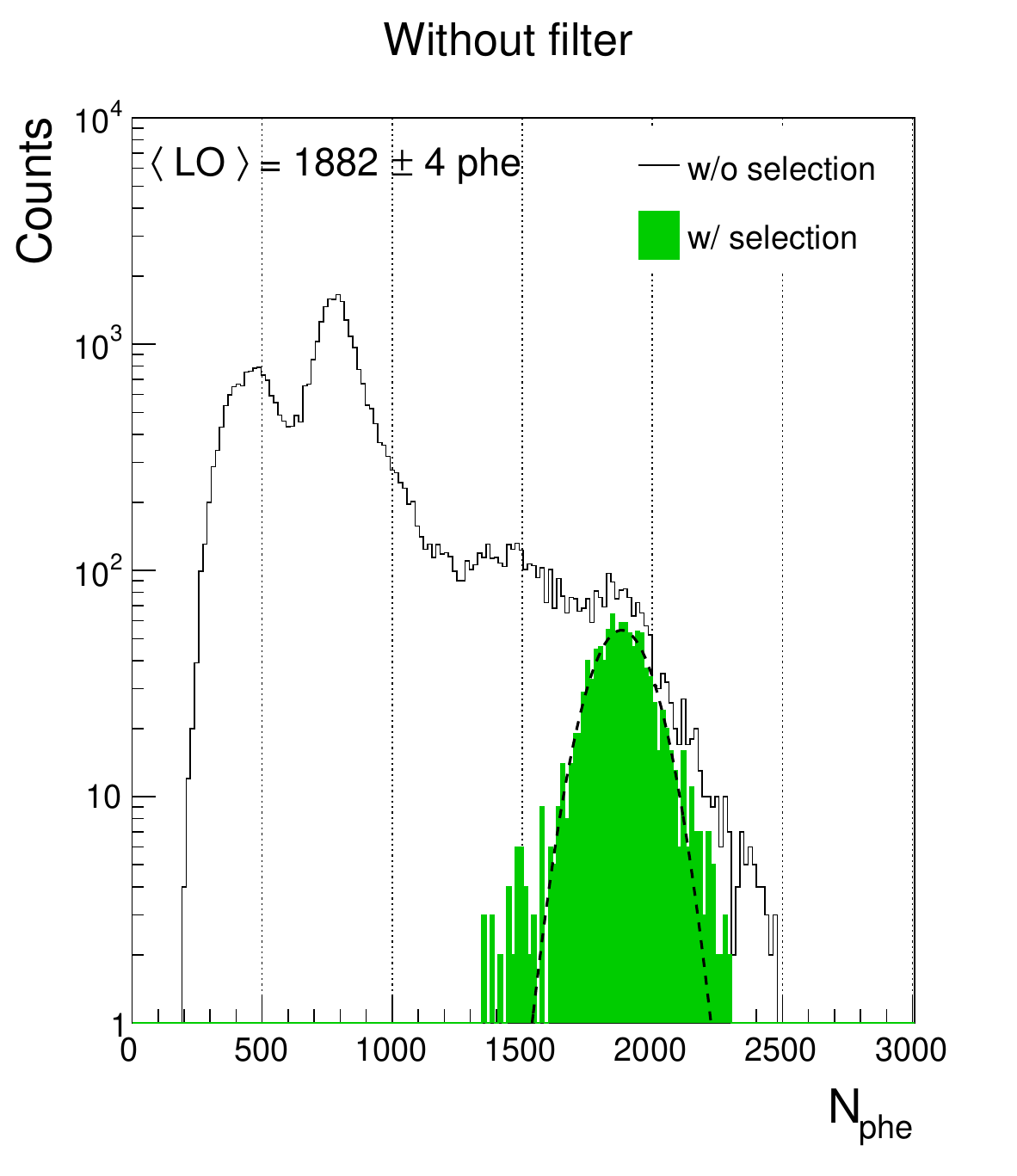}
    \includegraphics[width=0.325\linewidth]{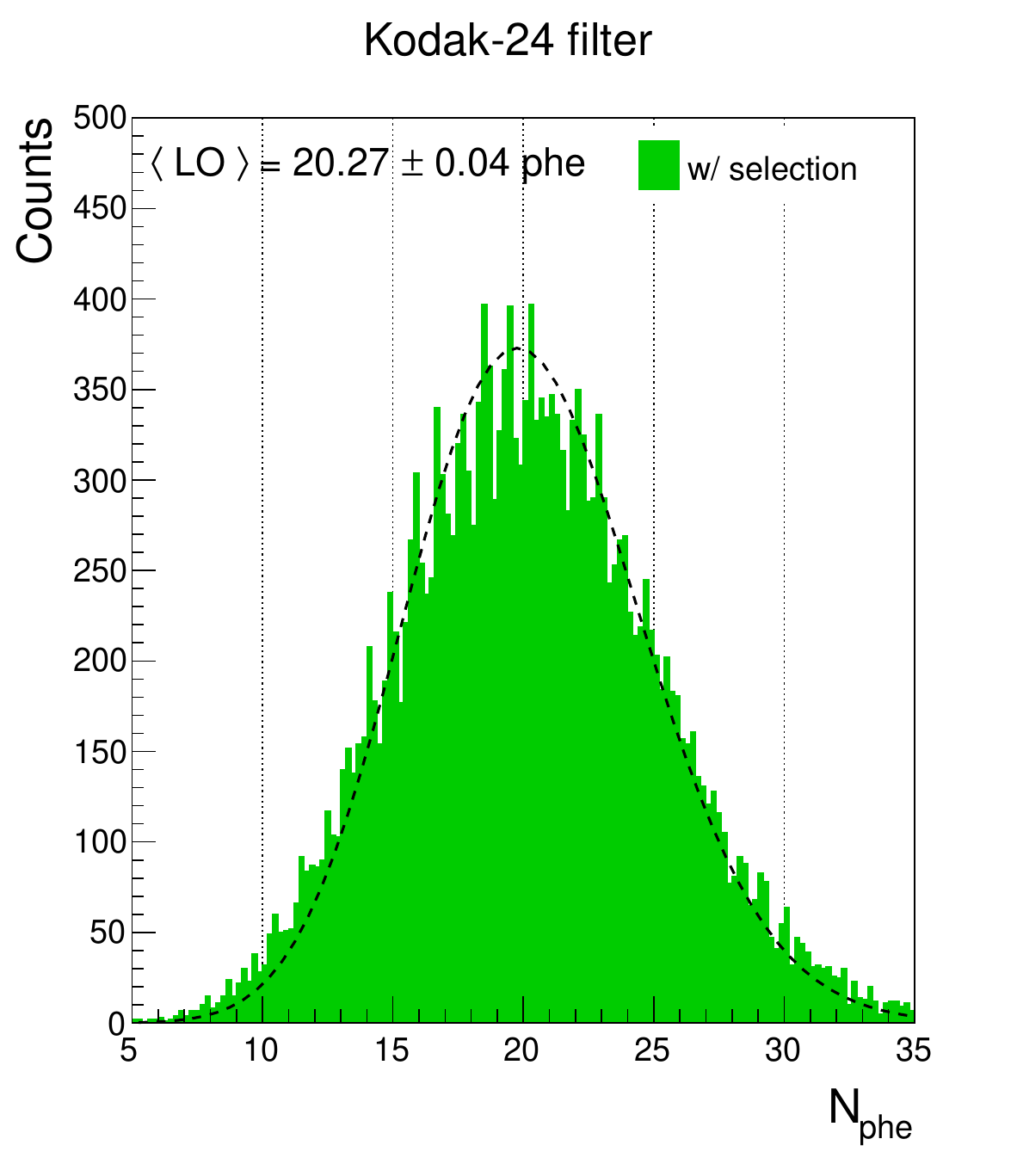}
    \includegraphics[width=0.325\linewidth]{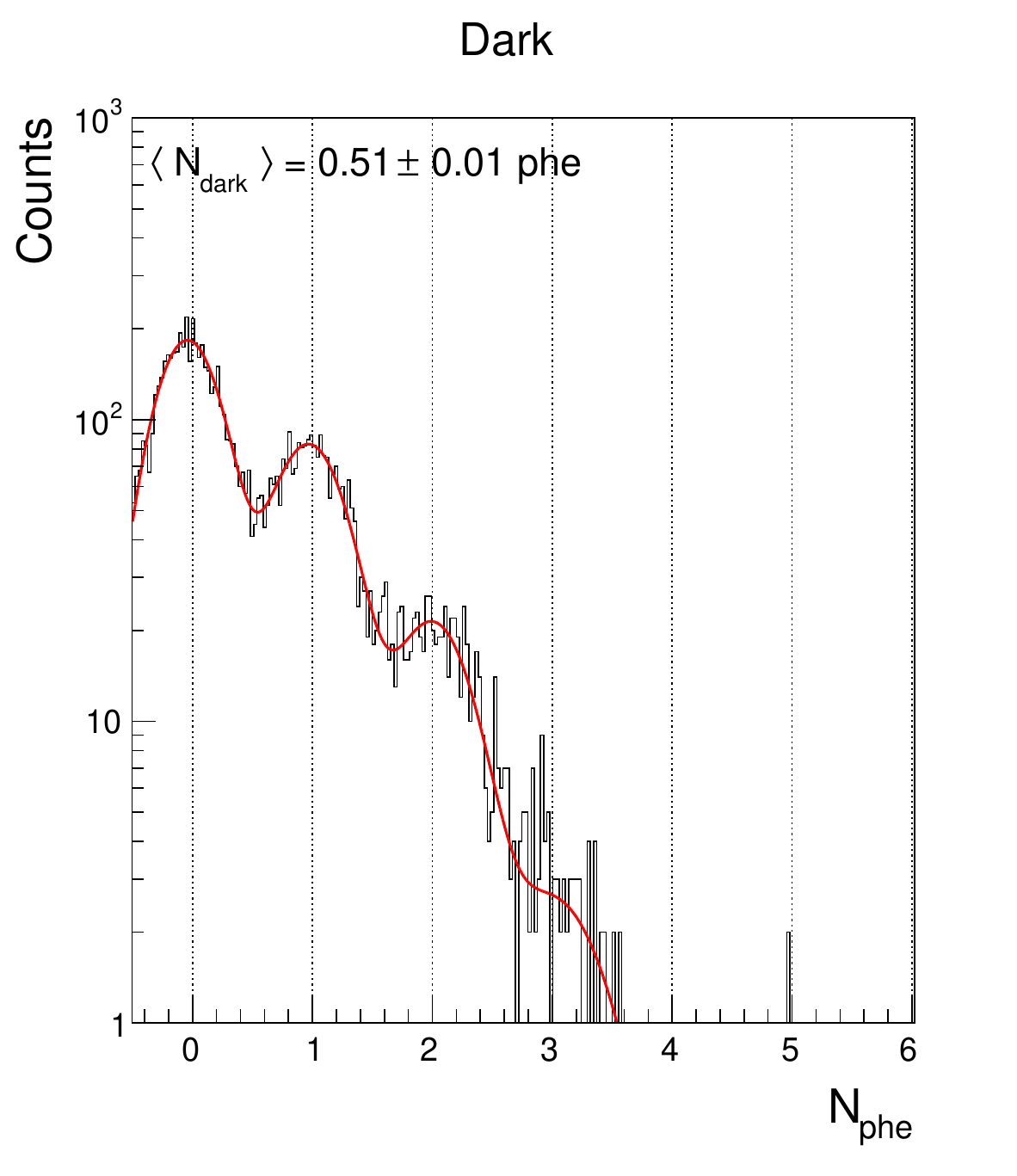}
    \caption{Left: spectrum of the LYSO:Ce crystal light output, calibrated in number of photoelectrons, for the main SiPM without filter with and without selection of events corresponding to the 1.275 MeV photopeak. Histogram with selected events is fit with a Gaussian function. Middle: Light output measured with a Kodak-24 filter between the crystal and the SiPM fitted with a Poisson continuous function. Right: spectrum of dark counts measured by shielding the main SiPM from the crystal and other light sources, and fitted with a sum of four Gaussian functions.}
    \label{fig:spectra}
\end{figure}

Events in which a 1275~keV $\gamma$-ray deposited its full energy via the photoelectric effect were selected using the reference SiPM, in order to maximize the light signal and thus the sensitivity of the measurement. The performance of each filter was then evaluated by computing the ratio of the measured light output with and without the filter at the main SiPM.
An example of typical spectra measured by the main SiPM is shown in Figure~\ref{fig:spectra} without filter (left panel) and with a Kodak-24 filter (central panel).
Without filter a light output of about 1480~phe/MeV (1888 phe at the 1275 ~keV photopeak) is measured while this drops to about 14~phe/MeV (20 phe at the 1275~keV photopeak) when a Kodak-24 filter is used. The contribution to the measured charge originating from SiPM dark counts that randomly occur within the signal integration window was estimated to be about 0.5~phe and is subtracted from the measured light outputs. Furthermore a correction to account for the SiPM saturation effect was applied as described in \cite{DRenker_2009}. The magnitude of this correction is about 5\% for the measurements without filter and negligible for the measurements with filter. As an example the light output without filter after correction was estimated to be 1560~phe/MeV (instead of 1480~phe/MeV).

A measurement using a double layer of Kodak-24 filters was also performed to assess whether a thicker filter could further improve the performance. A further reduction in the number of photoelectrons (from 20 to 15 at the 1275~keV peak) was observed, corresponding to a 25\% relative decrease (from 1.1\% to 0.8\%) in the transmitted scintillation light.
For comparison, an additional measurement was carried out using optical grease, yielding about 2700~phe/MeV (a factor of $\sim1.73$ higher) without the filter and 24~phe/MeV (a factor of $\sim1.71$ higher) with the Kodak-24 filter. These results indicate that the presence of optical grease (refractive index 1.55) between the crystal and the filter, and between the filter and the SiPM, does not significantly affect the filter performance.

\begin{figure}[!tbp]
    \centering
    \includegraphics[width=0.325\linewidth]{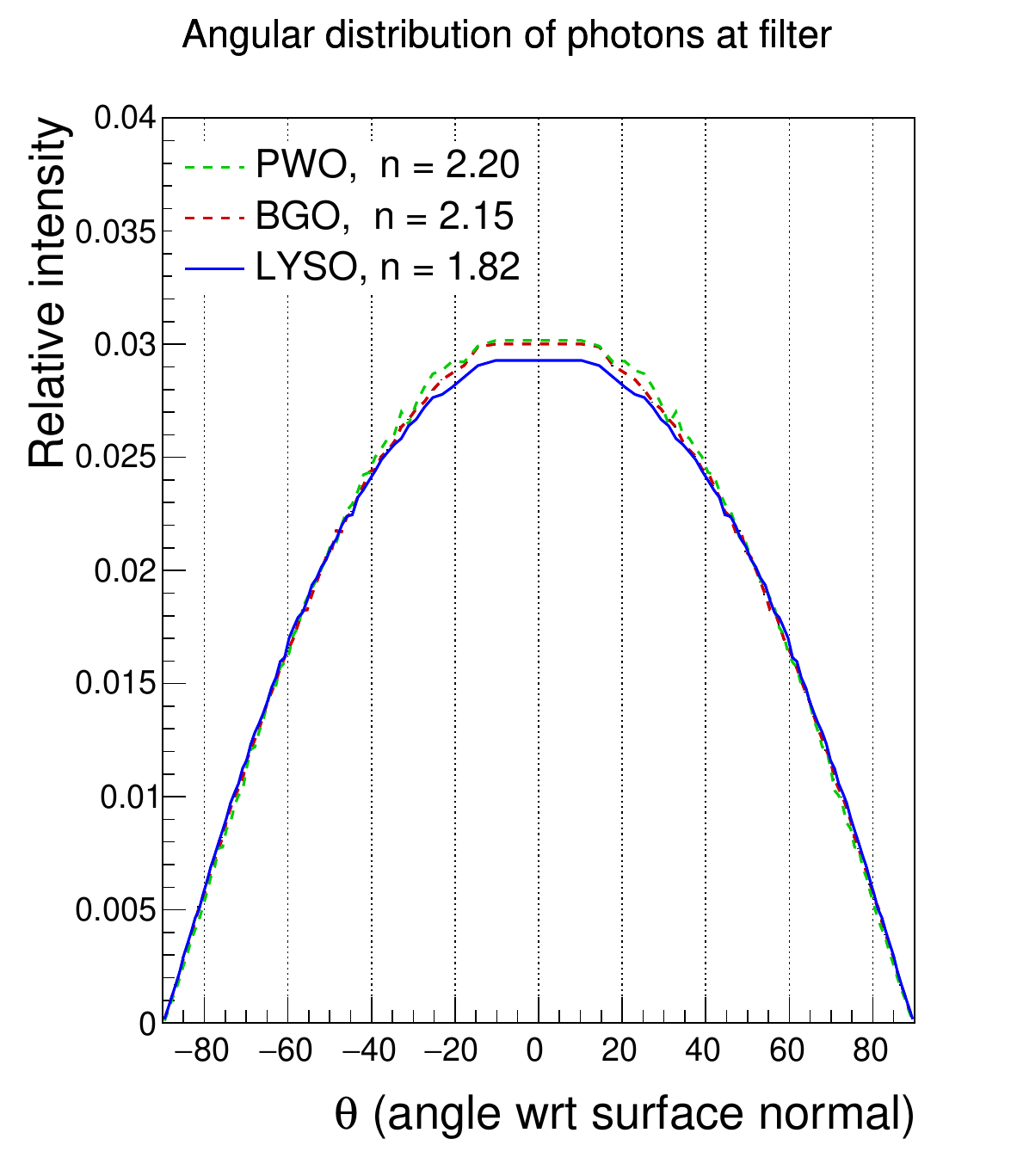}
    \includegraphics[width=0.325\linewidth]{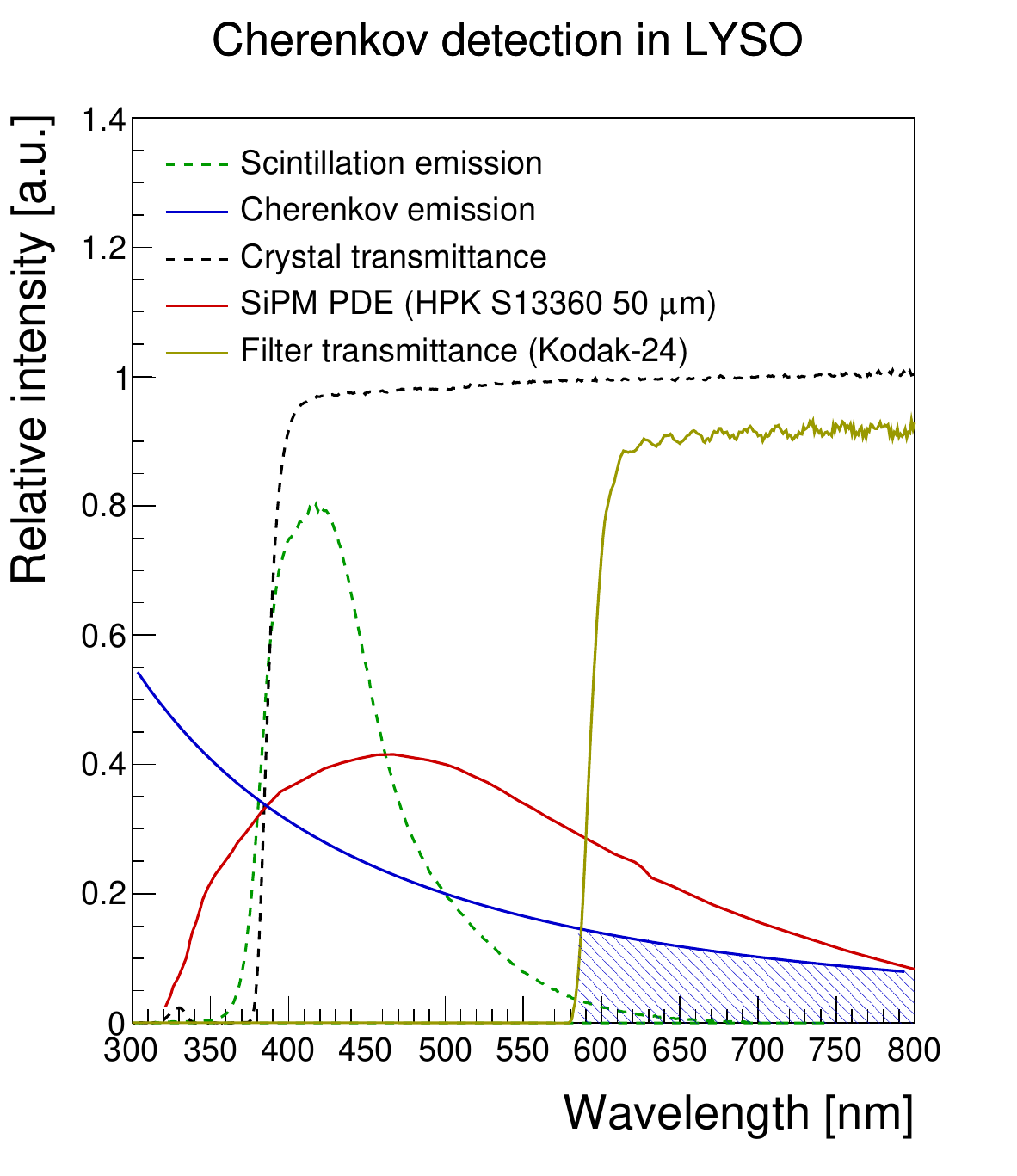}
    \includegraphics[width=0.325\linewidth]{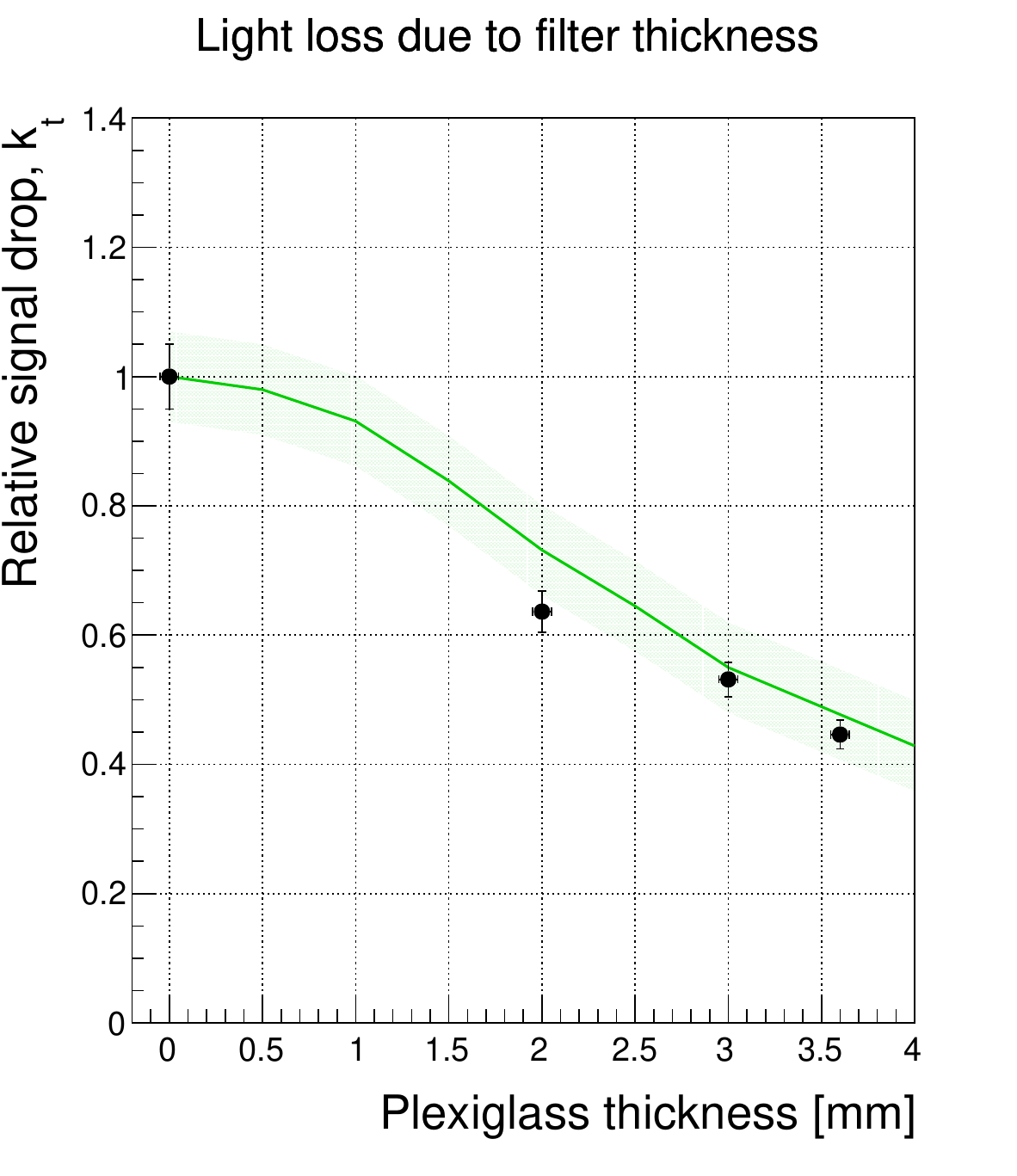}
    \caption{Left: photon angular distribution at the exit of the crystal (angle with respect to the normal to the filter surface). Middle: illustration of scintillation and Cherenkov photons emission wavelength compared to the crystal and (Kodak-24) filter transmittance. Right: measured drop in light output depending on the thickness of a transparent Plexiglass sample (black dots) compared to Geant4 simulation prediction (green band).}
    \label{fig:miscellanea}
\end{figure}

The results were then compared with calculations based on the crystal, filter and SiPM properties. Compared to the simple convolution of the emission spectrum with the filter transmittance in equation.~\ref{eq:scinti_fit}, the following additional effects must be taken into account:
\begin{itemize}
    \item {\bf Photon angular distribution}: The optical photons exit the crystal with a broad angular distribution due to the refractive index mismatch between the crystal and air. Figure~\ref{fig:miscellanea} shows the angular distribution of photons emerging from the crystal surface for LYSO, PWO, and BGO crystals, as simulated with Geant4. For interference filters, whose transmittance depends on the angle of incidence, this effect has been taken into account.
    \item {\bf SiPM Photon Detection Efficiency (PDE)}: The measured light output must also account for the photon detection efficiency (PDE) of the SiPM, $PDE(\lambda)$, which depends on the wavelength distribution of the photons incident on the SiPM. This distribution changes when an optical filter is applied. The photon detection efficiency of the SiPM used in this measurement is shown in Figure~\ref{fig:miscellanea}.
    \item {\bf Cherenkov photons background}: electrons produced by the photoelectric effect of 1.275~MeV $\gamma$-rays in LYSO:Ce have a mean range of 1.20~mm and can therefore generate Cherenkov photons along their tracks. These photons are emitted in random directions on average, since the initial electron direction is random and the track undergoes scattering at each interaction. Our Geant4 simulation predicts that, on average, 51 Cherenkov photons are produced in the wavelength range 350–1000~nm, following the usual $1/\lambda^2$ distribution (Figure~\ref{fig:miscellanea}). Of these, about 15 photons exit the crystal end face and, when accounting for the SiPM PDE, about 4.5 photoelectrons are detected without filter and 0.5–1.5 with filter. %This additional photon contribution, with a wavelength spectrum different from that of the scintillation light, has been included in the analysis.%, although it corresponds to only  additional photoelectrons depending on the filter.
    \item {\bf Filter thickness}: due to the large photon angular distribution, the presence of a thick filter between the crystal and the SiPM can lead to a loss of detected photons due to Fresnel refraction and pure geometrical effects. This effect, which is non-negligible for the Hoya filters, has been simulated with Geant4 and evaluated experimentally by measuring the relative drop in light output using a set of $\sim 1\rm~cm^2$ cross section samples of Plexiglass fully transparent to scintillation emission and with variable thickness. Results are shown in the right panel of Figure~\ref{fig:miscellanea} and are used to calculate a correction factor, $k_t$, in Eq.~\ref{eq:lo_ratio}.
\end{itemize}

\begin{figure}[!tbp]
    \centering
    \includegraphics[width=0.495\linewidth]{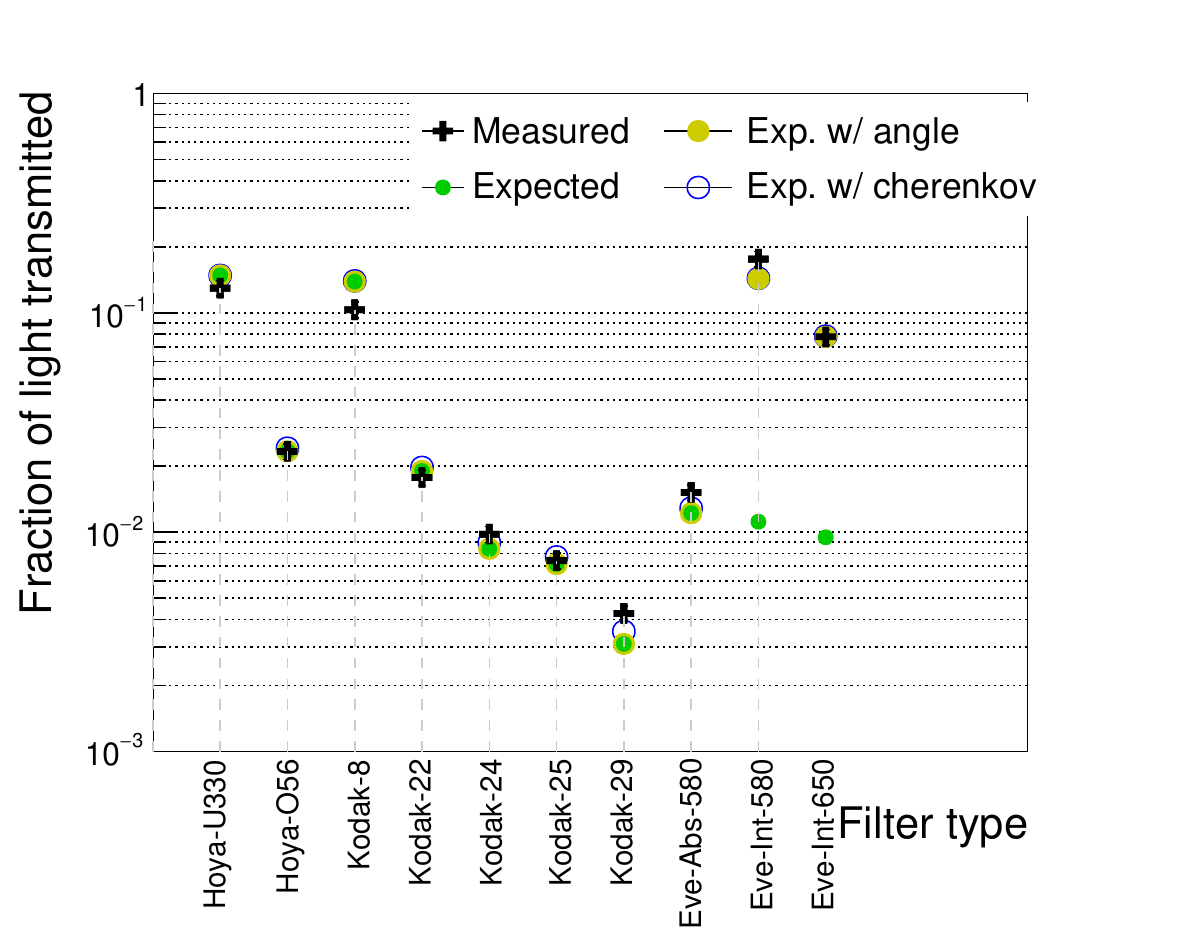}
    \includegraphics[width=0.495\linewidth]{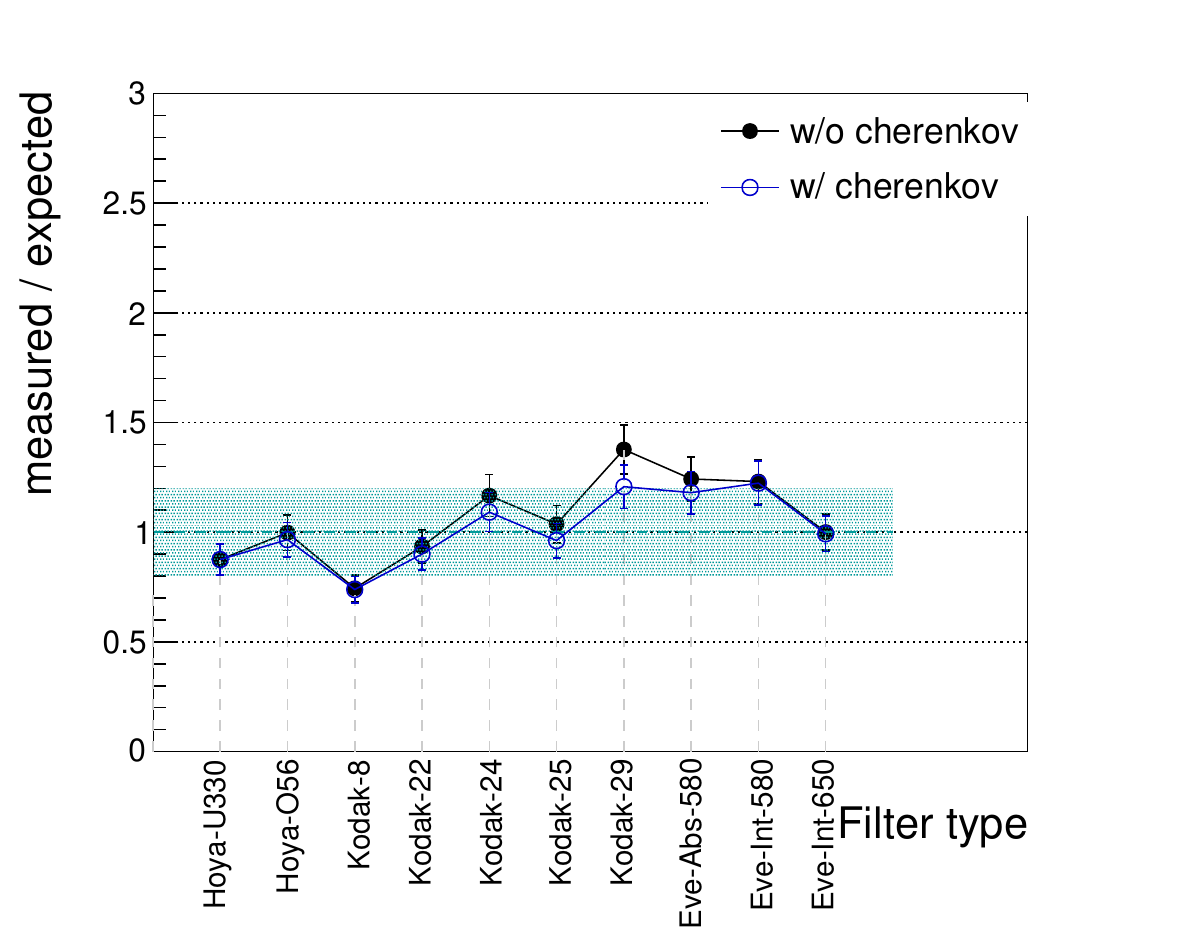}
    \caption{Left: Fraction of light output measured (black crosses) compared with the expected value at normal incidence (green dots), expected value accounting for filter angular dependence (yellow dot) and expected value when the contribution from Cherenkov photons is also included (blue circles). Right: ratio between the measured and expected fraction of light output with (black dots) and without (blue circles) accounting for the contribution from Cherenkov photons. The blue band indicates a 20\% uncertainty.}
    \label{fig:lo_ratio_source}
\end{figure}

\clearpage
\noindent
Considering all these aspects, we can calculate the expected ratio between the light output measured with and without filter as:
\begin{equation}\label{eq:lo_ratio}
    \rm \frac{LO_{filter}}{LO} = \frac{k_t\int_{-\pi/2}^{\pi/2}\int_{300~nm}^{800~nm} [(\alpha_s EM(\lambda) + \alpha_c C(\lambda)] \cdot T_c(\lambda)\cdot PDE(\lambda) \cdot {\bf T_f(\lambda, \theta) \xi(\theta)d\theta}  d\lambda }{\int_{300~nm}^{800~nm} [(\alpha_s EM(\lambda) + \alpha_c C(\lambda)] \cdot T_c(\lambda) \cdot PDE(\lambda) d\lambda }
\end{equation}
A comparison of the experimental results with calculations is shown in Figure~\ref{fig:lo_ratio_source}.
%The ratio between the measured and expected light output loss is shown in the right panel of Figure~\ref{fig:lo_ratio_source}. 

The agreement between the measured and expected fraction of light output is overall satisfactory, considering that the actual number of photons measured spans almost two orders of magnitude (from 0.4 to 20\%).
It is clear that the angular dependence of the interference filter must be taken into account to reproduce the measured values. This also implies that such filters are not effective at shielding optical photons emitted at broad angles from the crystal.
It can also be noted that accounting for Cherenkov photons becomes important when the number of detected scintillation photons is very small. For example, with the Kodak-29 filter, a total of about 10 photons is measured, of which roughly 1.5 are Cherenkov photons.%

\begin{figure}[!tbp]
    \centering
    \includegraphics[width=0.495\linewidth]{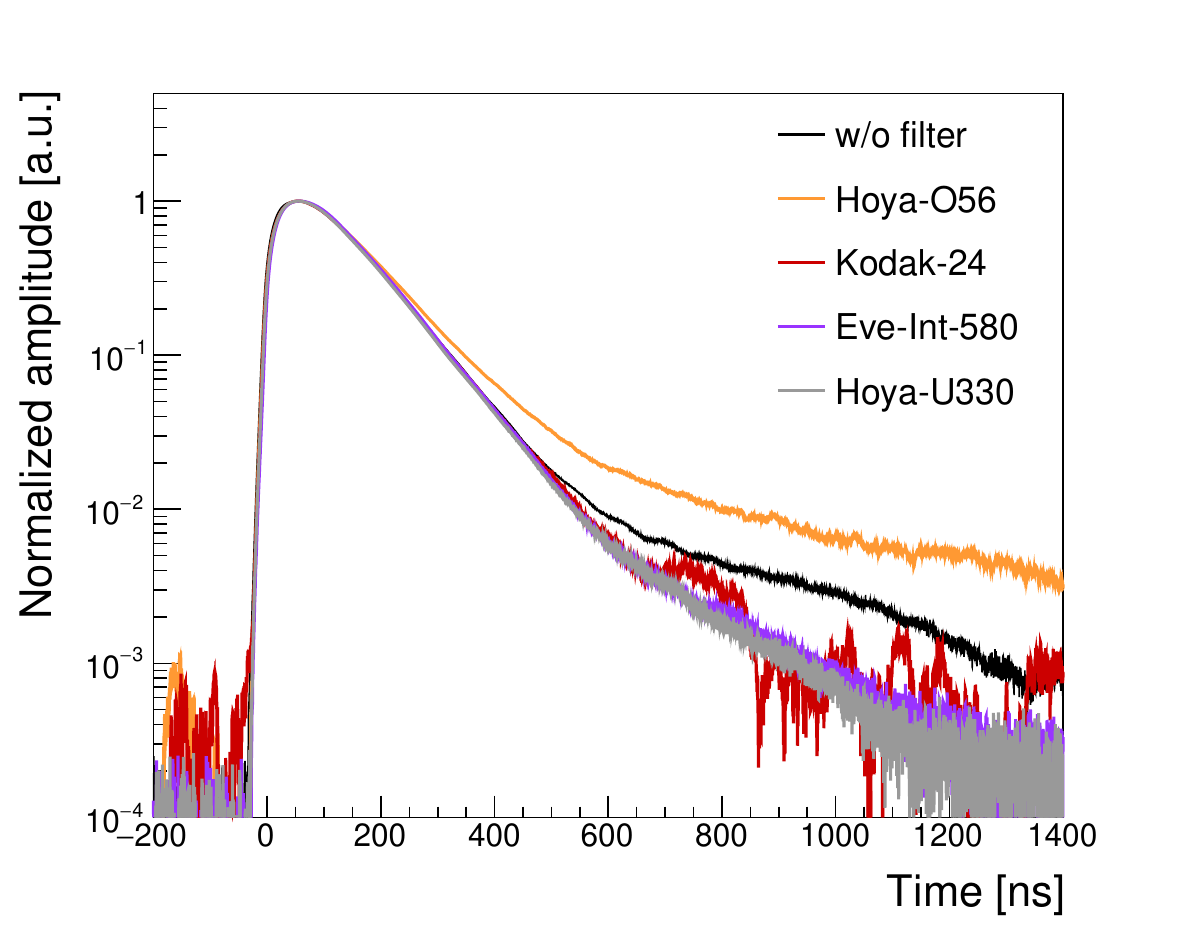}
    \includegraphics[width=0.495\linewidth]{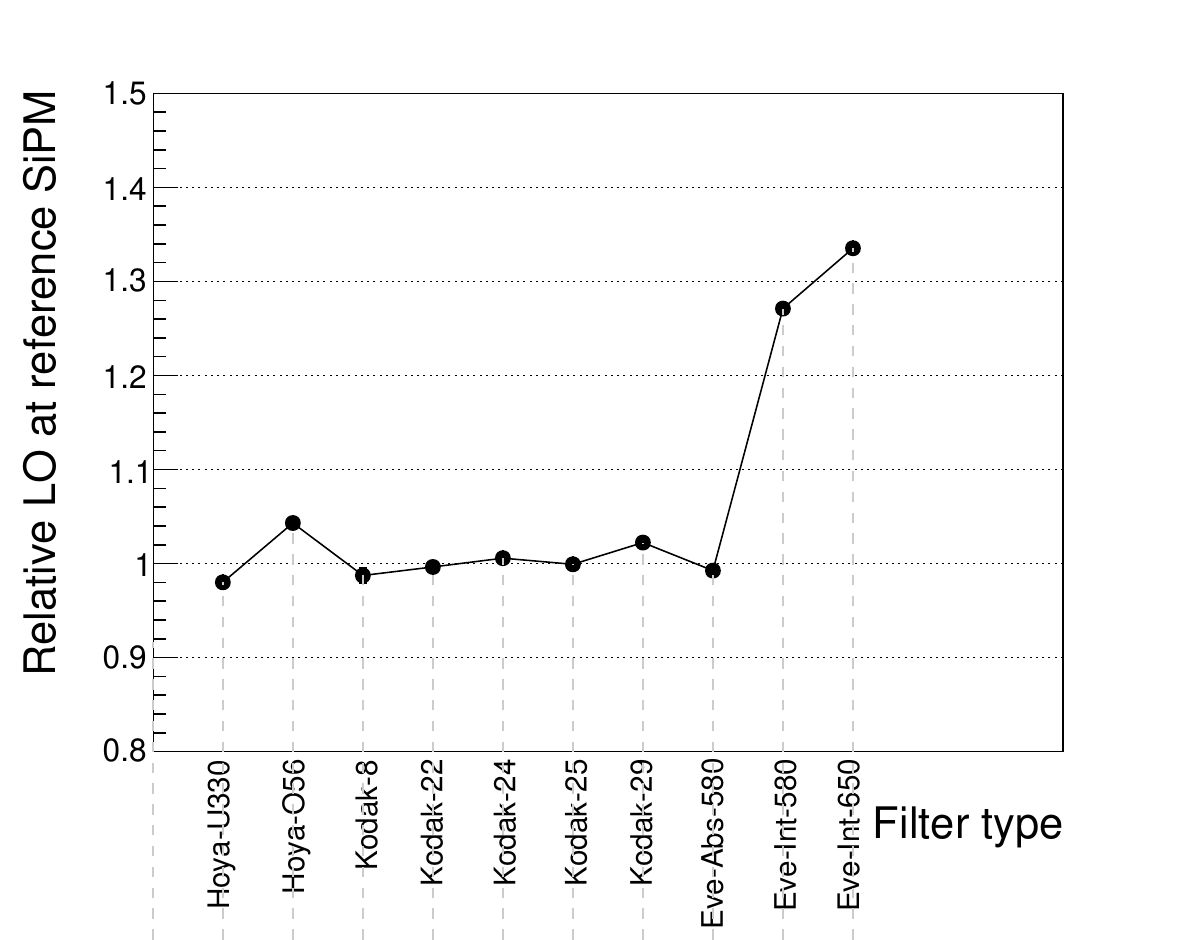}
    \caption{Left: Average waveforms recorded from the $3\times3\times5$~mm$^3$ LYSO:Ce crystal under $^{22}$Na source excitation with and without optical filters. The Hoya-O56 filter is the only that shows a slower signal suggesting the presence of delayed photons due to fluorescence. Right: relative light output measured at the reference SiPM while changing the filter at the opposite side of the crystal.}
    \label{fig:waveforms}
\end{figure}

Two additional observations were made during the measurements. First of all, from a comparison of the measured waveforms with and without filters, it emerged that only when the Hoya-O56 filter is used the average waveform features a longer decay time, as shown in the left panel of Figure~\ref{fig:waveforms}. This can be attributed to delayed photons originating from absorption of scintillation photons and re-emission of delayed photons at a longer wavelength by the filter itself.
The second observation is that when the two interference filters were tested a systematic net increase of the light output at the reference SiPM occurred, as shown in the right panel of Figure~\ref{fig:waveforms}.
This suggests that such filters, contrarily to absorptive ones, actually act as mirrors, reflecting inside the crystal a substantial fraction of the photons.

%\section{Modeling, performance extrapolations and outlook}

\section{Conclusions}
A set of scintillating crystals and optical filters representing technological candidates for a novel dual-readout electromagnetic calorimeter concept were studied in this work.
The performance of the filters in shielding the SiPM from scintillation light emitted by BGO, BSO, PWO, and LYSO crystals was calculated based on the measured crystal emission spectra and filter transmittance. The filter performance was also evaluated by directly measuring the light output from a LYSO crystal sample, with and without optical filters. Comparison of the measurements with expectations shows good agreement within 20\%.

The measurements indicate that interference (dichroic) filters are not suitable for this application, as their transmittance is angular dependent and does not effectively filter photons emitted at large angles from the crystal. Among the absorptive filters tested, the Hoya-O56 shows signs of delayed fluorescence, i.e. re-emission of scintillation photons at longer wavelengths \cite{filter_fluorescence}, which represents an unwanted feature for calorimetry applications. Furthermore, thick filters cause an overall wavelength-independent reduction of light output due to geometrical effects that also negatively affects the amount of Cherenkov photons collected at the SiPM.

Among the 100 \textmu m filters tested, the Kodak-24 and Kodak-25 samples, with cut-off wavelengths around 590 nm, show good performance for PWO crystals, filtering out more than 99\% of the light. The Kodak-29 sample, with a cut-off wavelength at 620 nm, can be used to block more than 97\% of the scintillation photons from BGO and BSO crystals, provided that additional discrimination between scintillation and Cherenkov photons based on their time distribution is applied.

\acknowledgments
This work received funding from the European Union in the framework of the Next Generation EU program, Mission 4, Component 1, CUP H53D23001120006. Part of this work was carried out in the frame work on Crystal Clear Collaboration and  received support from the Horizon Europe Twinning project TWISMA (GA 101078960).

%“Finanziato dall’Unione europea- Next Generation EU, Missione 4 Componente 1 CUP____________”.

\bibliography{mybibfile}

@ARTICLE{pagano_2022_decay,
AUTHOR={Pagano, Fiammetta  and Kratochwil, Nicolaus  and Frank, Isabel  and Gundacker, Stefan  and Paganoni, Marco  and Pizzichemi, Marco  and Salomoni, Matteo  and Auffray, Etiennette },
TITLE={A new method to characterize low stopping power and ultra-fast scintillators using pulsed X-rays},
JOURNAL={Frontiers in Physics},
VOLUME={10},
YEAR={2022},
URL={https://www.frontiersin.org/journals/physics/articles/10.3389/fphy.2022.1021787},
DOI={10.3389/fphy.2022.1021787},
ISSN={2296-424X}
}

@inproceedings{filter_fluorescence,
author = {S. Reichel and R. Biert{\"u}mpfel and A. Engel},
title = {{Characterization and measurement results of fluorescence in absorption optical filter glass}},
volume = {9626},
booktitle = {Optical Systems Design 2015: Optical Design and Engineering VI},
editor = {Laurent Mazuray and Rolf Wartmann and Andrew P. Wood},
organization = {International Society for Optics and Photonics},
publisher = {SPIE},
pages = {96260S},
year = {2015},
doi = {10.1117/12.2191047},
URL = {https://doi.org/10.1117/12.2191047}
}

@book{European:2720131,
      author        = "{The European Strategy Group}",
      title         = "{Deliberation document on the 2020 Update of the European Strategy for Particle Physics}",
      publisher     = "CERN Council",
      address       = "Geneva",
      year          = "2020",
      reportNumber  = "CERN-ESU-014",
      url           = "http://cds.cern.ch/record/2720131",
      doi           = "10.17181/ESU2020Deliberation",
}

@article{DRenker_2009,
doi = {10.1088/1748-0221/4/04/P04004},
url = {https://doi.org/10.1088/1748-0221/4/04/P04004},
year = {2009},
month = {apr},
volume = {4},
number = {04},
pages = {P04004},
author = {D Renker and E Lorenz},
title = {Advances in solid state photon detectors},
journal = {Journal of Instrumentation}
}

@article{Bacon_2025,
doi = {10.1088/1748-0221/20/09/P09012},
url = {https://doi.org/10.1088/1748-0221/20/09/P09012},
year = {2025},
month = {sep},
publisher = {IOP Publishing},
volume = {20},
number = {09},
pages = {P09012},
author = {Bacon, A. and Harris, B. and Klein, J.R.},
title = {Dichroic filter characterizations},
journal = {Journal of Instrumentation}
}

@article{RevModPhys.90.025002,
  title = {Dual-readout calorimetry},
  author = {Lee, Sehwook and Livan, Michele and Wigmans, Richard},
  journal = {Rev. Mod. Phys.},
  volume = {90},
  issue = {2},
  pages = {025002},
  numpages = {40},
  year = {2018},
  month = {Apr},
  publisher = {American Physical Society},
  doi = {10.1103/RevModPhys.90.025002},
  url = {https://link.aps.org/doi/10.1103/RevModPhys.90.025002}
}

@article{Lucchini_2020,
	url = {https://doi.org/10.1088/1748-0221/15/11/p11005},
	year = 2020,
	month = {nov},
	publisher = {{IOP} Publishing},
	volume = {15},
	number = {\textbf{11}},
	pages = {P11005--P11005},
	author = {M.T. Lucchini and W. Chung and S.C. Eno and Y. Lai and L. Lucchini and M. Nguyen and C.G. Tully},
	title = {New perspectives on segmented crystal calorimeters for future colliders},
	journal = {Journal of Instrumentation}
}

@article{Ampilogov_2023,
doi = {10.1088/1748-0221/18/09/P09021},
url = {https://dx.doi.org/10.1088/1748-0221/18/09/P09021},
year = {2023},
month = {sep},
publisher = {IOP Publishing},
volume = {18},
number = {09},
pages = {P09021},
author = {N. Ampilogov and S. Cometti and J. Agarwala and V. Chmill and R. Ferrari and G. Gaudio and P. Giacomelli and A. Giaz and A. Karadzhinova-Ferrer and A. Loeschcke-Centeno and A. Negri and L. Pezzotti and G. Polesello and E. Proserpio and A. Ribon and R. Santoro and I. Vivarelli},
title = {Exposing a fibre-based dual-readout calorimeter to a positron beam},
journal = {Journal of Instrumentation}
}

@article{Addesa_2022,
doi = {10.1088/1748-0221/17/08/P08028},
url = {https://dx.doi.org/10.1088/1748-0221/17/08/P08028},
year = {2022},
month = {aug},
publisher = {IOP Publishing},
volume = {17},
number = {08},
pages = {P08028},
author = {F.M. Addesa and P. Barria and R. Bianco and M. Campana and F. Cavallari and A. Cemmi and M. Cipriani and I. Dafinei and B. D'Orsi and D. del Re and M. Diemoz and G. D'Imperio and E. Di Marco and I. Di Sarcina and M. Enculescu and E. Longo and M.T. Lucchini and F. Marchegiani and P. Meridiani and S. Nisi and G. Organtini and F. Pandolfi and R. Paramatti and V. Pettinacci and C. Quaranta and S. Rahatlou and C. Rovelli and F. Santanastasio and L. Soffi and R. Tramontano and C.G. Tully},
title = {Comparative characterization study of LYSO:Ce crystals for timing applications},
journal = {Journal of Instrumentation}
}

\end{document}